\makeatletter\def\input@path{{main_support/}}\makeatother
\documentclass{optica-article}

\journal{opticajournal}
\articletype{Research Article}

\hypersetup{pdfauthor={Sui Ip Ng, Yi-Ping Ho, Ken-Tye Yong}}
\graphicspath{{figures/}}

\begin{document}

\title{Axial von Mises Framework for Tensor-Dependent Second-Harmonic Generation}

\author{Sui Ip NG\authormark{1}, Yi-Ping Ho\authormark{1,2}, and Ken-Tye YONG\authormark{3,*}}

\address{\authormark{1}Department of Biomedical Engineering, The Chinese University of Hong Kong, Hong Kong SAR, China\\
\authormark{2}Department of Mechanical Engineering, National Taiwan University, Taiwan\\
\authormark{3}School of Biomedical Engineering, Faculty of Engineering, The University of Sydney, Sydney, NSW 2008, Australia}

\email{\authormark{*}ken.yong@sydney.edu.au}

\begin{abstract*}
We derive closed-form relations between projected fibril-axis statistics and tensor-dependent second-harmonic generation (SHG) harmonics using a single axial von Mises population. Under coherent emission within domains and detected intensity addition across domains, each angular harmonic factors into a local tensor coefficient and an axial circular moment. Parallel-analyzer detection contains second, fourth and sixth harmonics; ideal equal-weight analyzer-free detection contains second and fourth harmonics. The resulting formulas identify geometry-specific harmonic zeros, sensitivity loss and ambiguous inversion of a secondary fixed-angle ratio. In particular, the analyzer-free fourth-harmonic zero at $\rho=1/3$ does not imply loss of all orientation sensitivity. Prepared-cornea forward/reverse recordings illustrate extraction of empirical analyzer-free harmonics. In the bundle ROI, second harmonics are resolved in both sweeps, while their normalized amplitudes differ and fourth harmonics remain unresolved. An exploratory central ROI from the same recordings yields a different forward/reverse profile. Within the same operator-identified tissue, low-signal space coverage and apparent fibre alignment vary slightly across recordings. These observations support angular-measurement feasibility while leaving quantitative normalized-shape agreement unestablished. The thermal extension illustrates consequences of assumed kinetics rather than calibrated tissue trajectories. The contribution is an analytical interpretation of SHG harmonics with explicit conditions for physical inference; the measurements do not establish absolute susceptibility ratio $\rho$, axial concentration $\kappa$, or experimental separation of thermal disorder from degradation.
\end{abstract*}

\section{Introduction}

Second-harmonic generation (SHG) microscopy provides angularly dependent contrast in collagen-rich tissues \cite{campagnola2002three,williams2005interpreting,stoller2002polarization}. Interpreting this contrast requires distinguishing the local nonlinear susceptibility from the distribution of projected fibril axes and the detected intensity scale. Polarization-resolved measurements have been used to characterize tissue organization and susceptibility ratios \cite{PLOTNIKOV2006693,tiaho2007estimation}, but neither an angular ratio nor a change in total intensity is a universal measure of orientational order.

Here we focus on a compact analytical description with explicit structural and detection assumptions. Its interpretation requires specifying which orientation statistics enter each angular harmonic, how they combine with the local tensor response, and when a normalized metric loses sensitivity or admits more than one inverse solution. The following section places this objective within established polarization-resolved SHG methods and defines the scope of the present study.

\section{Prior Work and Study Objective}
\label{sec:prior_work}

\subsection{Established P-SHG Approaches}

Established P-SHG methods extract collagen orientation and information about the nonlinear optical response. Stoller et al.\ demonstrated polarization modulation measurements of fiber orientation and a parameter related to the nonlinear susceptibility, while noting its sensitivity to disorganization, oblique incidence and birefringence \cite{stoller2002polarization}. Orientation contrast and interpretation informed by the nonlinear tensor are therefore established P-SHG capabilities.

Hierarchical and disordered-bundle models are also established. Tuer et al.\ combined polarization-in/polarization-out measurements with a hierarchical model relating fibril architecture to effective susceptibility ratios \cite{tuer2012hierarchical}. Rou\`ede et al.\ explicitly modeled disordered fibrils within a three-dimensional cone, together with helical hierarchy and photonic shot noise, and used linear least-squares analysis of polarization stacks to interpret several tissues \cite{rouede2017determination}. Their model is a relevant alternative, not a special case automatically validated by the present two-dimensional intensity average. It constructs effective tensor components by summing fibril contributions, whereas we add detected domain intensities after local field contraction. Rou\`ede's effective $\rho=\chi_{33}/\chi_{31}$ also differs in definition and structural level from our local $\rho=\chi_{uvv}/\chi_{uuu}$. Published values cannot be transferred by tissue name or by taking a reciprocal alone.

Distribution-based and efficient full-scan analyses precede this work. Duboisset et al.\ introduced a generic approach for retrieving mean molecular orientation and accessible symmetry orders without prescribing an orientational distribution family \cite{duboisset2012generic}. Radaelli et al.\ used coupled phasor projections of polarization stacks to retrieve orientation and anisotropy without direct nonlinear fitting \cite{radaelli2017mmapps}. Fast estimation and Fourier-based compression of P-SHG data are therefore not novel capabilities claimed here. Compared with these approaches, our restriction to one axial von Mises population sacrifices distributional flexibility in exchange for explicit Bessel-moment relations and interpretation of their tensor-dependent sensitivity and inverse branches.

Axial von Mises statistics are themselves established in corneal SHG. Raoux et al.\ fitted lamellar-orientation distributions with two axial von Mises populations and an isotropic component \cite{raoux2023unveiling}; von Mises fits to image-derived collagen orientations also appear in the mitral-valve study of Sadeghinia et al.\ \cite{sadeghinia2024planar}. Such image-derived fiber statistics, molecular orientational orders and projected domain-axis statistics describe different structural levels. Their concentration parameters are not interchangeable without a specified mapping. The present use of an axial von Mises family is a declared model choice, not a claim to introduce that family into collagen imaging.

Thermal structural assessment is likewise an established application. Matteini et al.\ used SHG polarization anisotropy to investigate changes in interfibrillar organization and molecular structure during corneal heating \cite{matteini2012thermal}. Consequently, sensitivity to thermal change is not, by itself, the distinguishing contribution sought here. The thermal example in this paper illustrates consequences of the specified harmonic model under assumed kinetics, rather than establishing an experimentally validated early-damage detector.

The physical averaging rule is a further boundary. Polarimetric calculations by Alizadeh et al.\ show that fiber polarity, crossing and out-of-plane configuration can affect intensity and inferred susceptibility ratios when field amplitudes and phase relations are considered \cite{alizadeh2023partially}. The domain-intensity average adopted here excludes detected cross-domain interference. Its conservation of angular mean under pure orientation redistribution is therefore conditional on that assumption and fixed response/weighting; it is not a universal law of collagen SHG or an established separation of disorder from degradation.

\subsection{Study Objective and Scope}

Our goal is \textbf{a compact, tensor-aware axial model that provides explicit sensitivity limits and interpretation rules for SHG angular harmonics}.

The results provide three linked interpretation tools: explicit tensor-factor/circular-moment expressions for each detectable harmonic; geometry-specific conditions for harmonic zeros and sensitivity loss; and calibration/inverse-branch criteria for interpreting normalized angular observables. Within the chosen von Mises family, the circular moments are Bessel-function ratios controlled by $\kappa$. This restricted family makes the relations explicit, while the separate parallel-analyzer and ideal equal-weight analyzer-free contractions prevent transfer of a criterion between incompatible observables. The experimental assessment uses empirical harmonic coefficients, independently of fitting physical $\rho$ or $\kappa$, so the illustrative optical evidence is not used circularly to calibrate the model parameters.

The framework applies to full angular scans; the two-angle ratio is a secondary summary. We do not establish that it outperforms established full-scan fitting in parameter accuracy, photon efficiency or acquisition speed. Its possible reduced-acquisition use is conditional on prior calibration and stability, rather than a replacement for the information needed to establish those conditions.

The thermal calculation serves as an illustrative extension, and the prepared-cornea measurements provide a preliminary analyzer-free feasibility and repeatability assessment. The current experiment does not establish absolute $\rho$ or $\kappa$, the exact $\rho=1/3$ crossover, or thermal disorder/degradation separation. Neither improved early thermal detection performance nor a unique orientation family is inferred from the pilot.

\section{Theory}
\label{sec:theory}

\subsection{Von Mises Model and RA-SHG Observables}

We model the angular distribution of fibril axes using the axial von Mises distribution on the circle \cite{mardia2009directional, benlakhdar2024depth, landler2018circular},
\begin{equation}
f_{\rm ax}(\psi\mid\mu,\kappa)=
\frac{\exp[\kappa\cos2(\psi-\mu)]}{2\pi I_0(\kappa)},
\label{eq:vm}
\end{equation}
where $\kappa\geq0$ is the concentration parameter ($\kappa\gg1$: tightly aligned; $\kappa\to0$: isotropic). The physical fibril-axis angle is $\psi$, and $f_{\rm ax}(\psi+\pi)=f_{\rm ax}(\psi)$. The density is normalized over a full $2\pi$ interval; over an interval of length $\pi$, the denominator is $\pi I_0(\kappa)$. The distribution describes the \textit{projected} orientation in the 2D imaging plane; out-of-plane orientation requires a separate projection and optical-response model. The axial moments are $M_n(\kappa)=\langle\cos[2n(\psi-\mu)]\rangle=I_n(\kappa)/I_0(\kappa)$, with vanishing sine moments \cite{mardia2009directional, hillen2017moments}. Odd physical-angle moments, including $\langle\cos\psi\rangle$ and $\langle\cos^3\psi\rangle$, vanish over the full circle. For large $\kappa$, the local angular variance is approximately $1/(4\kappa)$.

\begin{figure}[htbp]
\centering
\includegraphics[width=0.9\textwidth]{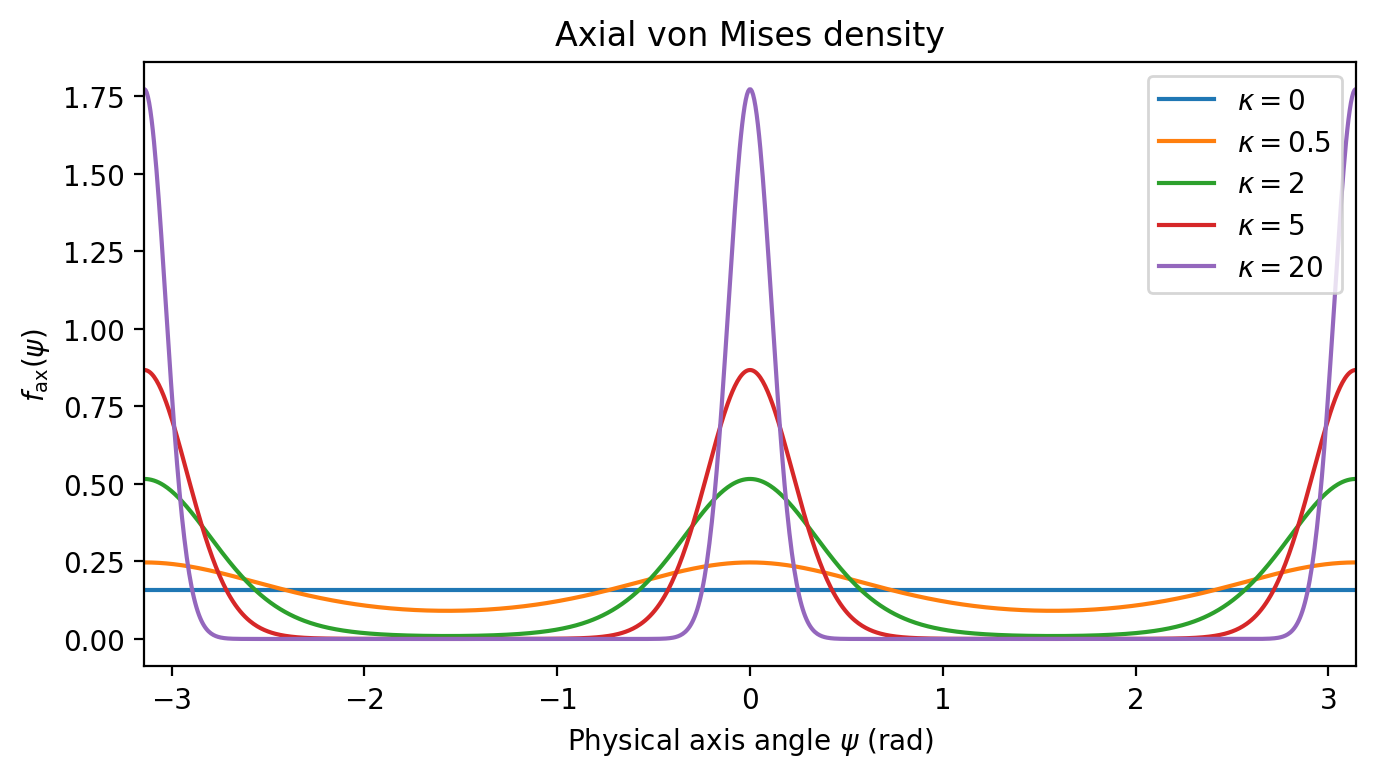}
\caption{Axial von Mises distributions for the concentration values shown. The two equivalent axis directions differ by $180^\circ$. Concentration describes projected axis statistics within the assumed family; it is not an absolute parameter measured by the present pilot.}
\label{fig:axial_distribution}
\end{figure}

We first consider the rank-one tensor limit ($\rho=0$) to isolate the orientational degree of freedom. We assume coherent emission within each domain but negligible detected interference between domains, so domain intensities are added. This assumption must be justified experimentally; axial symmetry alone does not establish incoherent addition. Averaging the signed rank-three tensor over equal opposite polarities would instead give zero coherent mean field \cite{oudar1977hyperpolarizabilities, boyd2023nonlinear}. For incident field $\mathbf{E}(\omega)=E_0(\cos\phi,\sin\phi,0)$ and parallel analyzer, intensity averaging gives, with $\mu=0$ and a common scale removed,
\begin{equation}
I_\parallel(\phi;\kappa)=
\int_0^{2\pi} f_{\rm ax}(\psi\mid0,\kappa)\cos^6(\phi-\psi)\,d\psi .
\label{eq:I_explicit}
\end{equation}
This expands as $I_\parallel(\phi)=c_0+c_2\cos2\phi+c_4\cos4\phi+c_6\cos6\phi$, with
\begin{equation}
c_0=\frac{10}{32},\qquad
c_2=\frac{15M_1}{32},\qquad
c_4=\frac{6M_2}{32},\qquad
c_6=\frac{M_3}{32}.
\label{eq:coeffs_A}
\end{equation}

We define the fixed-angle ratio $C_{\rm raw}\equiv I_\parallel(0^\circ)/I_\parallel(45^\circ)$, with both angles relative to the mean axis. For $\rho=0$, $I_\parallel(0^\circ)=(10+15M_1+6M_2+M_3)/32$ and $I_\parallel(45^\circ)=(10-6M_2)/32$, giving
\begin{equation}
C_{\rm raw}(\kappa)=
\frac{10+15M_1+6M_2+M_3}{10-6M_2}.
\label{eq:contrast_raw}
\end{equation}
The limits are $C_{\rm raw}\to8$ (aligned, $\kappa\to\infty$) and $C_{\rm raw}\to1$ (isotropic, $\kappa\to0$). Normalizing,
\begin{equation}
\tilde{C}(\kappa)\equiv\frac{C_{\rm raw}-1}{7},
\label{eq:contrast_norm}
\end{equation}
maps $[1,8]\to[0,1]$ for $\rho=0$. This normalization uses the rank-one reference limits; the mapping between $\tilde{C}$ and $\kappa$ depends on $\rho$ (Section~\ref{sec:finite_rho}), so $\tilde{C}$ is a tensor-dependent fixed-angle ratio rather than a universal contrast metric. Negative values indicate ratios below the isotropic reference of one, not negative intensity. The polarization-averaged signal is $S\propto\beta^2c_0(\rho)$, where $\beta$ is a common molecular-response amplitude. In this intensity-averaged model, $c_0$ is independent of $\kappa$.

\subsection{Extension to Finite \texorpdfstring{$\rho$}{rho}}
\label{sec:finite_rho}

We now incorporate the in-plane collagen susceptibility tensor under cylindrical and Kleinman symmetry \cite{PLOTNIKOV2006693, boyd2023nonlinear, chu2004studies}:
\begin{equation}
\chi_{ijk}(\psi)=\chi_1u_i u_j u_k+
\chi_2(u_i v_j v_k+v_i u_j v_k+v_i v_j u_k),
\label{eq:chi_local}
\end{equation}
with $\mathbf{u}=(\cos\psi,\sin\psi,0)$, $\mathbf{v}=(-\sin\psi,\cos\psi,0)$, and real $\rho\equiv\chi_2/\chi_1\geq0$. The local laboratory-frame components are
\begin{align}
\chi_{xxx}(\psi)&=\chi_1[\cos^3\psi+3\rho\cos\psi\sin^2\psi],
\label{eq:chi_xxx_rho}\\
\chi_{xyy}(\psi)&=\chi_1[\cos\psi\sin^2\psi+
\rho(\cos^3\psi-2\cos\psi\sin^2\psi)].
\label{eq:chi_xyy_rho}
\end{align}
Their signed axial averages vanish; the detected observable is obtained by averaging squared local fields instead.

Writing $\delta=\phi-\psi$, the normalized local parallel field is
$g_\rho(\delta)=P\cos^3\delta+Q\cos\delta$, where
\begin{equation}
P(\rho)=1-3\rho,\qquad Q(\rho)=3\rho,\qquad
g_\rho(\delta)=\cos^3\delta+3\rho\cos\delta\sin^2\delta.
\label{eq:PQ_def}
\end{equation}
Here $P,Q$ are local-field coefficients independent of $\kappa$, rather than components of an axially averaged macroscopic susceptibility. The intensity
$I_\parallel=\int f_{\rm ax}(\psi\mid\mu,\kappa)g_\rho(\phi-\psi)^2\,d\psi$ admits the Fourier expansion
\begin{equation}
I_\parallel(\phi;\kappa,\rho)=c_0+
c_2\cos2(\phi-\mu)+c_4\cos4(\phi-\mu)+c_6\cos6(\phi-\mu).
\label{eq:I_fourier_rho}
\end{equation}
Its closed-form coefficients (see Supplement 1 for full derivation) are
\begin{equation}
\begin{aligned}
c_0&=\frac{5P^2}{16}+\frac{3PQ}{4}+\frac{Q^2}{2}\equiv d_0,\\
c_2&=\left(\frac{15P^2}{32}+PQ+\frac{Q^2}{2}\right)M_1\equiv d_2M_1,\\
c_4&=\left(\frac{3P^2}{16}+\frac{PQ}{4}\right)M_2\equiv d_4M_2,\qquad
c_6=\frac{P^2}{32}M_3\equiv d_6M_3.
\end{aligned}
\label{eq:coeffs_rho}
\end{equation}

The factorization follows from an angular convolution. For either local intensity kernel, write $q(\delta)=q_0+\sum_{n\geq1}q_{2n}\cos2n\delta$, with the finite harmonic orders fixed by the detection contraction. The axial distribution gives
\[
\int_0^{2\pi} f_{\rm ax}(\psi\mid\mu,\kappa)
\cos[2n(\phi-\psi)]\,d\psi
=M_n(\kappa)\cos[2n(\phi-\mu)].
\]
Thus orientation averaging multiplies each local harmonic by its corresponding circular moment and leaves the constant term $q_0$ unchanged. The orientation-independent mean is a consequence of a normalized angular average of a fixed local kernel. It requires the stated intensity-addition rule and fixed tensor/collection weighting; it is not a physical conservation law established by the cornea recordings.
The fixed-angle ratio becomes
\begin{equation}
C_{\rm raw}(\kappa,\rho)=
\frac{d_0+d_2M_1+d_4M_2+d_6M_3}{d_0-d_4M_2},
\label{eq:contrast_raw_rho}
\end{equation}
with $\tilde{C}=(C_{\rm raw}-1)/7$. For $\rho=0$, this reduces to Eq.~\eqref{eq:contrast_norm}. For finite $\rho$, $C_{\rm raw}(0,\rho)=1$ and $C_{\rm raw}(\infty,\rho)=8/(1+3\rho)^2$. The perfectly aligned state has $\tilde{C}<0$ for $\rho>(\sqrt8-1)/3$, but the ratio need not remain negative or monotone for all finite $\kappa$. Its interpretation requires a calibrated $\rho$ and, where necessary, an independently justified inverse branch.

\subsection{Remarks on the Choice of Angles}

$C_{\rm raw}=I_\parallel(0^\circ)/I_\parallel(45^\circ)$ is a \textit{fixed-angle ratio}, not a true contrast. The sampled angles are relative to the mean axis and need not coincide with angular extrema. Unlike the coherently averaged polar model, $I_\parallel(90^\circ)$ is generally nonzero at finite $\kappa$. The finite ratio at $(0^\circ,45^\circ)$ provides a practical experimental metric, but tensor-dependent off-axis maxima and nonmonotonicity require full-scan checks.

Two measurements do not retain the independent harmonic amplitudes, phase information or residual diagnostics of a full scan. They cannot determine an unknown mean axis, tensor ratio, concentration and background together. Computing a ratio from an acquired full scan adds no information; normalized harmonic analysis already removes a common multiplicative scale. The ratio's possible benefit is fewer subsequent acquisitions after the axis, tensor/background response and usable inverse branch have been established and shown to remain stable. Sequential gain drift, axis motion or changing $\rho$ invalidates that simplification. No universal advantage at fixed acquisition time or photon budget is demonstrated here. Full scans remain the appropriate starting point for calibration, model checking and joint uncertainty assessment.

With a common detected amplitude $G\beta^2$ and additive background $B$, the measured ratio is
\begin{equation}
\tilde{C}_{\rm meas}=
\frac{1}{7}\left[
\frac{G\beta^2I_\parallel(0^\circ)+B}
{G\beta^2I_\parallel(45^\circ)+B}-1\right].
\end{equation}
As $\kappa\to0$, both intensities approach $G\beta^2d_0+B$, giving $\tilde{C}_{\rm meas}\to0$. A fixed background suppresses anisotropy and couples the measured ratio to molecular signal loss; background subtraction is needed to recover the scale-independent ratio.

\subsection{Tensor-Dependent Endpoint and Monotonicity Criteria}
\label{sec:sign_inversion}

\textbf{Endpoint criterion.} For $\tilde{C}(\kappa,\rho)$ with $B=0$:
\begin{enumerate}
\item For $0\leq\rho<\rho_*=(\sqrt8-1)/3$, the aligned endpoint satisfies $\tilde{C}(\infty)>0$, whereas the isotropic endpoint is zero.
\item For $\rho>\rho_*$, the aligned endpoint satisfies $\tilde{C}(\infty)<0$, whereas the isotropic endpoint is zero.
\end{enumerate}
Equality of endpoint ratios at $\rho_* \simeq0.6095$ is not a global monotonicity criterion. In particular, at $\rho=1/3$ the local field is $g_\rho(\delta)=\cos\delta$, so
\begin{equation}
I_\parallel(\phi)=\frac12[1+M_1\cos2(\phi-\mu)],
\qquad C_{\rm raw}=1+M_1.
\end{equation}
The ratio remains order-sensitive, so there is no universal sign inversion at $\rho=1/3$ in this axial intensity-averaged model.

\textit{Reasoning.} The aligned ratio is $8/(1+3\rho)^2$, while the isotropic ratio is one. Near isotropy,
\begin{equation}
C_{\rm raw}=1+\frac{d_2}{2d_0}\kappa+O(\kappa^2),
\qquad d_2=\frac{3(1+\rho)(5-3\rho)}{32}.
\end{equation}
For $\rho_*<\rho<5/3$, the slope near isotropy is positive but the aligned ratio is below one. The mapping is therefore nonmonotone, including for $\rho\approx0.62$, and a ratio-only inverse can be ambiguous. Such an interior maximum can occur even when $B=0$ and cannot be attributed solely to background.

Physically, disorder attenuates the second, fourth, and sixth harmonics through different axial moments, while leaving $c_0$ fixed. Their tensor-dependent combination determines the fixed-angle ratio. Full polarization scans and independently calibrated $\rho$ are therefore needed to assess local identifiability; no universal native-collagen $\kappa$ range removes these ambiguities.

\begin{figure}[htbp]
\centering
\includegraphics[width=0.9\textwidth]{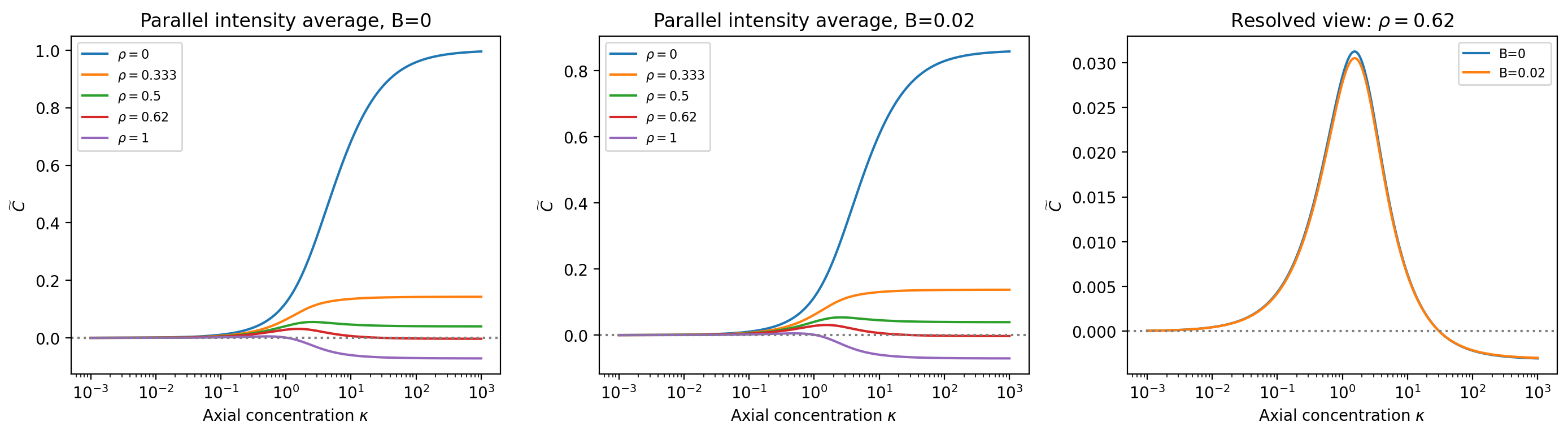}
\caption{Normalised fixed-angle ratio $\tilde{C}(\kappa,\rho)$ for several $\rho$. (a) Background-free and (b) with additive background $B=0.02$. The response depends on the tensor ratio and is not universally monotonic. (c) Resolved comparison for $\rho=0.62$: an interior maximum occurs even at $B=0$ and is therefore not solely a background artifact. Background suppresses the magnitude of the metric toward zero. Equality of the aligned and isotropic endpoints at $\rho_*=(\sqrt{8}-1)/3$ does not establish global monotonicity.}
\label{fig:finite_rho}
\end{figure}

\subsection{Analyzer-Free Harmonics}
\label{sec:AF}

Detection without an output polarization analyzer requires a different contraction from the parallel-analyzer observable. With equal detected weighting of the two in-plane components, adding their squared local fields gives
\begin{equation}
I_{\rm AF}(\phi)=D_0+D_2M_1\cos2(\phi-\mu)
+D_4M_2\cos4(\phi-\mu),
\label{eq:AF}
\end{equation}
where
\begin{equation}
D_0=\frac{3+2\rho+7\rho^2}{8},\qquad
D_2=\frac{1-\rho^2}{2},\qquad
D_4=\frac{(1+\rho)(1-3\rho)}{8}.
\label{eq:AFcoeffs}
\end{equation}
The derivation is given in Supplement 1, Section S2. There is no sixth harmonic in this ideal geometry. The background-subtracted harmonic ratios are $D_2M_1/D_0$ and $D_4M_2/D_0$ in the mean-axis frame; their interpretation depends on tensor ratio, background, collection weighting and the assumed orientation family.

For a single symmetric axial population with $\kappa>0$, the fourth harmonic changes sign at $\rho=1/3$. The second harmonic remains nonzero there, since $D_2=4/9$. This is a conditional analytical result and is distinct from the parallel-analyzer fixed-angle endpoint criterion. An unresolved fourth harmonic cannot locate a specimen at this boundary: weak order, measurement uncertainty, mixed orientations and optical response can also limit its identification. The present experiment does not test the exact crossover or fit the physical parameters in Eq.~\eqref{eq:AF}.

\begin{figure}[htbp]
\centering
\includegraphics[width=0.95\textwidth]{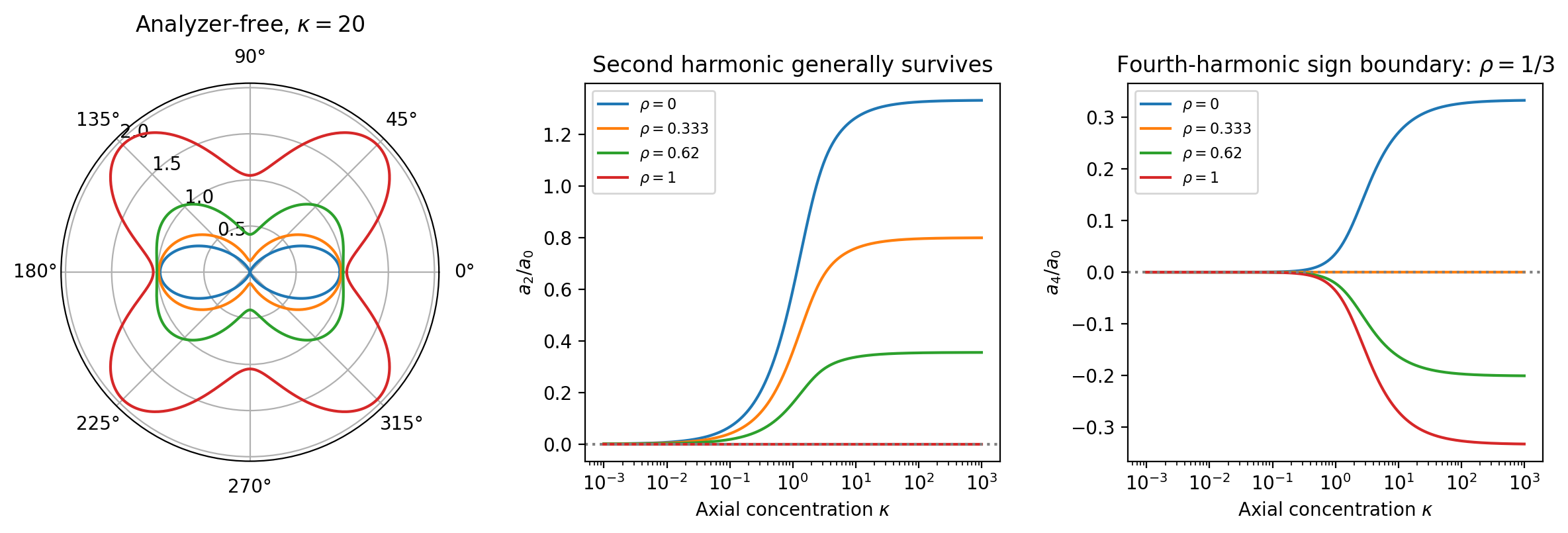}
\caption{Analyzer-free model calculations: angular patterns and normalized second/fourth harmonics. Equal in-plane collection weighting and domain-intensity addition are imposed. The fourth-harmonic zero at $\rho=1/3$ is a model property, not an experimentally established tissue boundary.}
\label{fig:AF}
\end{figure}

\section{Identifiability and Conditional Precision}
\label{sec:uncertainty}

Physical identifiability must be distinguished from numerical fit precision. For a measured response $Y(\phi)=G\beta^2I(\phi;\kappa,\rho,\mu)+B$, the transformation
\[
\beta\longmapsto a\beta,\qquad G\longmapsto G/a^2,
\qquad a>0,
\]
leaves every predicted intensity unchanged. Absolute molecular-response amplitude therefore requires independent information about the collection/excitation scale. Likewise, one fixed-angle scalar supplies only one relation between unknown $\kappa$ and $\rho$; fixing $\rho$ still does not remove the inverse-branch ambiguity of Section~\ref{sec:sign_inversion}. A calibrated full scan may constrain shape parameters under the assumed model, but neither a small residual nor a local precision bound removes these structural ambiguities.

\subsection{Prospective Concentration Inference}

The following analysis concerns prospective inference under calibrated, identifiable conditions; no $\kappa$ estimate is extracted from the experimental pilot. Even under these conditions, measurement noise limits precision. With independent, equal relative intensity uncertainties $\sigma_I/I$, the raw ratio $R\equiv I_0/I_{45}$ has $\sigma_R/R=\sqrt2\,\sigma_I/I$, and $\tilde{C}=(R-1)/7$ inherits
\begin{equation}
\sigma_{\tilde{C}}=\frac{1+7\tilde{C}}{7}\sqrt2\,\frac{\sigma_I}{I},
\qquad
\sigma_\kappa\simeq
\frac{\sigma_{\tilde{C}}}{|d\tilde{C}/d\kappa|}.
\end{equation}
These formulas require a specified background treatment and a locally identifiable inverse branch. For $U=d_0+d_2M_1+d_4M_2+d_6M_3$ and $V=d_0-d_4M_2$,
\begin{equation}
\frac{d\tilde{C}}{d\kappa}=\frac{U'V-UV'}{7V^2},\qquad
M_n'=\tfrac12(M_{n-1}+M_{n+1})-M_nM_1,\quad M_0=1 .
\end{equation}
Uncertainty becomes large where the ratio saturates or its derivative vanishes. Neither a universal $\kappa$ cutoff nor a universal SNR guarantee follows from the axial model. Uncertainty from $\rho$ and background calibration must also be propagated. The inverse bounds can be asymmetric, with their direction depending on the branch; we recommend reporting $\kappa^{+\Delta\kappa_+}_{-\Delta\kappa_-}$ from inversion of $\tilde{C}_{\rm meas}\pm2\sigma_{\tilde{C}}$ on a justified branch, or reporting all admissible intervals where multiple branches exist.

\begin{figure}[htbp]
\centering
\includegraphics[width=0.65\textwidth]{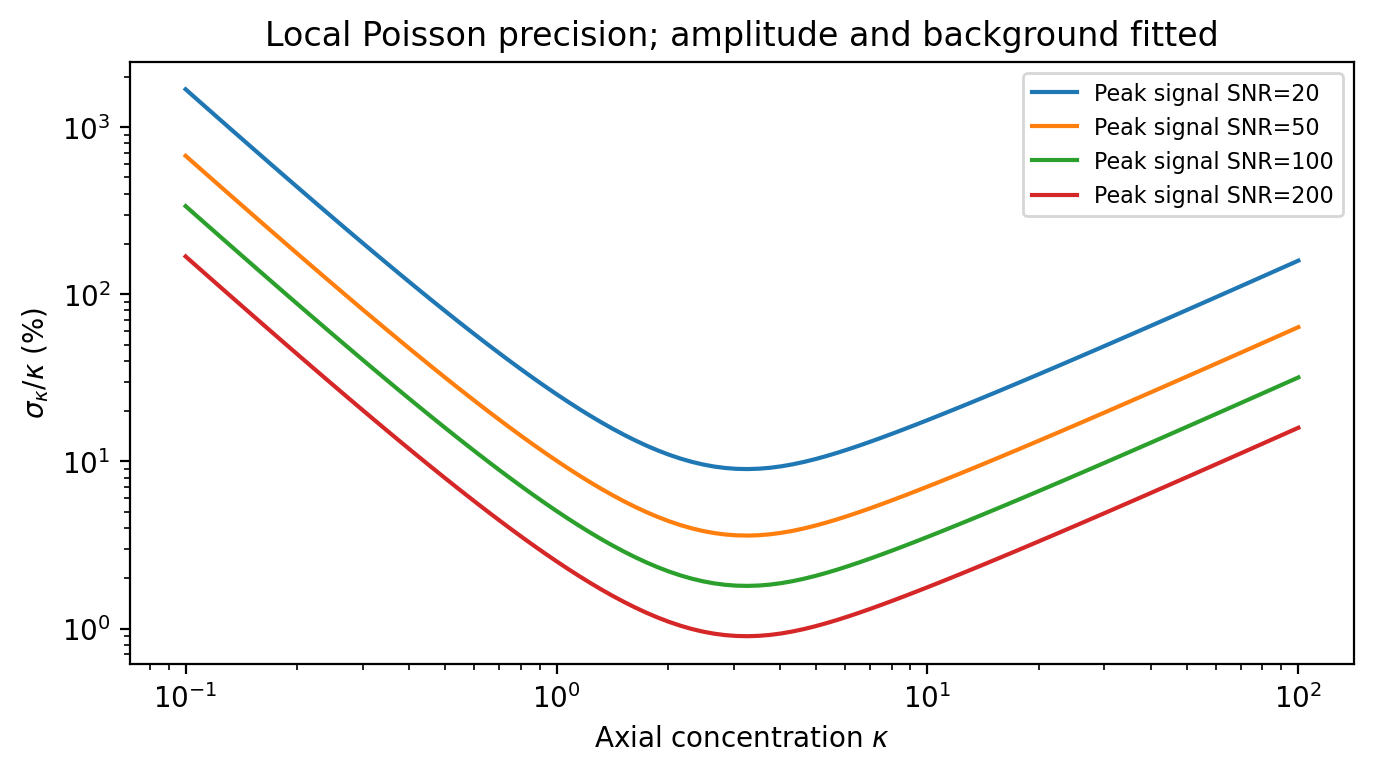}
\caption{Ideal local Poisson Fisher lower bound on relative concentration uncertainty for 36-angle parallel-analyzer scans, with $\rho=0$ and mean axis fixed, and amplitude and background treated as nuisance parameters. Peak signal SNR denotes $\sqrt{N_{\rm peak,sig}}$; background is $b=0.02A$ and scales with signal. Consequently, uncertainty scales inversely with this count parameter without changing the curve shape. These illustrative bounds exclude systematic errors and are not empirical fit uncertainties.}
\label{fig:kappa_uncertainty}
\end{figure}

The preceding propagation formulas apply to a two-angle ratio, whereas Fig.~\ref{fig:kappa_uncertainty} uses the information in a full parallel-analyzer scan. For independent Poisson counts at 36 equally spaced angles $\phi_i\in[0,\pi)$, let $\lambda_i=A I_\parallel(\phi_i;\kappa,\rho=0,\mu=0)+b$. The local Fisher matrix for $\boldsymbol{\eta}=(\kappa,\log A,\log b)$ is
\begin{equation}
F_{ab}=\sum_i\frac{1}{\lambda_i}
\frac{\partial\lambda_i}{\partial\eta_a}
\frac{\partial\lambda_i}{\partial\eta_b},
\qquad \sigma_{\kappa,{\rm F}}=\sqrt{(F^{-1})_{\kappa\kappa}} .
\end{equation}
Amplitude and background are nuisance parameters; the tensor ratio and mean axis are fixed. This is a local Cramer--Rao lower bound for regular unbiased estimation, not an uncertainty measured from repeated fits. Singular or poorly conditioned cases are excluded.

The plotted count parameter is $s=\sqrt{N_{\rm peak,sig}}$, labelled peak signal SNR, with $A=s^2/\max_i I_\parallel(\phi_i)$ and $b=0.02A$. It is not a background-inclusive measured SNR: at the peak, signal divided by total Poisson standard deviation is $N_{\rm peak,sig}/\sqrt{N_{\rm peak,sig}+b}$. Increasing $s$ scales both signal and background counts by the same factor. Thus $F\propto s^2$ and
\begin{equation}
\frac{\sigma_{\kappa,{\rm F}}}{\kappa}=\frac{h(\kappa)}{s},
\end{equation}
where $h$ includes angular sensitivity and nuisance-parameter correlations. The unchanged curve shape and constant vertical shifts on logarithmic axes therefore follow from the acquisition assumptions, rather than a universal experimental scaling law.

These bounds are optimistic and conditional on a correct, stable, shot-noise-only model. Fixed background, detector read noise, or fitting the mean axis and tensor ratio can alter the precision curves. Calibration errors, drift, and model mismatch may impose an uncertainty floor; near weak identifiability or parameter boundaries, fitted errors may be larger, biased, and asymmetric.

\subsection{Tensor Calibration, Bias and Validity Range}
\label{sec:rho_calibration}

The interpretation of $\kappa$ requires information about the local tensor ratio beyond a fitted fixed-angle contrast. A practical optical route is calibrated input/output polarimetry: characterize excitation and detection polarization transfer, acquire polarization-in/polarization-out data or resolved orthogonal SHG channels, and relate the recovered response to a local tensor using independently constrained orientation, symmetry and optical geometry \cite{dow2016unified,lee2023calibrated}. For dynamic measurements, simultaneous orthogonal-channel detection or calibrated rapid polarization modulation should resolve the relevant temporal changes; a prior unheated calibration cannot establish that $\rho$ remains fixed during heating. A reference crystal checks instrument response; it does not determine the specimen's $\rho$. Molecular tilt, chirality, mixed orientations and interference can still confound an effective fitted ratio \cite{alizadeh2023partially}. Thus fitting $\rho$ and $\kappa$ freely to the same scalar is not independent calibration, and no generally valid isolation of intrinsic $\rho$ in heterogeneous tissue is asserted here. When independent constraints are unavailable, report harmonic observables or admissible parameter sets instead of a unique concentration trajectory.

Let $F(\kappa,\rho)=\tilde C(\kappa,\rho)$ denote the background-free parallel-analyzer metric. If the true ratio differs from the value held fixed in inversion by $\delta\rho=\rho_{\rm true}-\rho_{\rm assumed}$, the local systematic bias is
\begin{equation}
\hat\kappa-\kappa\simeq
\frac{F_\rho}{F_\kappa}\,\delta\rho .
\label{eq:rho_bias}
\end{equation}
This first-order approximation fails near a stationary point or saturation, where the complete inverse set must be considered. Table~\ref{tab:review_sensitivity} gives finite, noise-free inversions for the requested $\pm10\%$ perturbation. For nominal $\rho=0.2$ and true $\kappa=5$, the inferred values are approximately 6.999 and 3.876, biases of $+40.0\%$ and $-22.5\%$. A 10\% ratio error therefore does not imply a 10\% concentration error. For nominal $\rho=0.62$, true $\kappa=5$ and true $\rho$ 10\% higher, two solutions occur, approximately 0.0158 and 25.063. Neither is selected as a preferred branch. Supplement 1, Section S10, reports the full computational range, ambiguous cases and failures of admissible inversion. These are illustrative model checks, not tissue parameter estimates or experimental calibration.

\begin{table}[htbp]
\centering
\caption{Illustrative inferred $\hat\kappa$ at nominal $\rho=0.2$. Tensor checks use true $\rho=0.18$ or 0.22 with zero background. The background check instead uses the correct $\rho=0.2$ and an unsubtracted background-to-signal ratio $B/S_{45}=0.1$. An em dash denotes no solution in the computational search $0\leq\hat\kappa\leq1000$; this is not a biological bound.}
\label{tab:review_sensitivity}
\begin{tabular}{rrrr}
\hline
True $\kappa$ & $\rho_{\rm true}=0.18$ & $\rho_{\rm true}=0.22$ & $B/S_{45}=0.1$\\
\hline
2 & 2.183 & 1.839 & 1.743 \\
5 & 6.999 & 3.876 & 3.747 \\
20 & \textemdash & 8.295 & 8.656 \\
\hline
\end{tabular}

\end{table}

We do not prescribe $T<60^\circ$C as a universally valid range. Corneal structural changes and tendon signal changes depend on the treatment protocol, and changes have been reported below that proposed cutoff \cite{matteini2012thermal,sun2006investigating}. The applicability criterion is stability or independent characterization of the tensor response, orientation/collection model and background over the observation interval. Heating through helix unwinding or a symmetry change without such information invalidates a fixed-$\rho$ inversion, irrespective of the nominal cutoff. Below $60^\circ$C is not automatically validated, and the present unheated pilot establishes no thermal operating range.

\subsection{Background Controls and Concentration Inference}
\label{sec:background_controls}

A deterministic additive background must be distinguished from zero-mean measurement noise. Write the true signal at $45^\circ$ as $S_{45}=G\beta^2I_\parallel(45^\circ)$, and let $\widehat B$ be the estimated common background. For residual $e=(B-\widehat B)/S_{45}$, the corrected metric satisfies the exact identity
\begin{equation}
\tilde C_{\rm corrected}=\frac{F(\kappa,\rho)}{1+e},
\qquad 1+e>0.
\label{eq:background_residual}
\end{equation}
Ignoring the background gives $e=B/S_{45}>0$; excessive subtraction gives $e<0$. The local concentration bias is $\hat\kappa-\kappa\simeq-F e/F_\kappa$ for small $e$ on a regular branch. As signal changes, a fixed physical background produces a varying $e(T)$ and can mimic an order change. At nominal $\rho=0.2$, $B/S_{45}=0.1$ changes an inferred true concentration of 5 to approximately 3.747; for a true value of 20 it gives approximately 8.656. A common background fraction is therefore not a universal concentration-error tolerance. Supplement 1, Section S10, shows the resulting $\hat\kappa(T)$ curves, errors from 10\% background misestimation, and a constant-order synthetic control.

For prospective physical-parameter inference, the following controls are required:
\begin{enumerate}
\item Measure detector dark/offset records and an appropriate illuminated blank under matched gain, exposure, excitation and optical settings. A dark frame does not measure illuminated stray light or specimen fluorescence. Retain any angle/time dependence of background; Eq.~\eqref{eq:background_residual} assumes one common background at the two angles.
\item Subtract the independently estimated background on the intensity scale, retaining its uncertainty and shared covariance. Preserve signed observations; do not clip negative corrected values or change ROIs to improve the recovered trajectory. An unmeasured specimen-dependent background remains an uncertainty, not a freely chosen correction.
\item Prefer an appropriate count or detector-noise likelihood for the original intensities when the corrected denominator is uncertain. Estimate background as a constrained nuisance quantity using the controls, and propagate its uncertainty through all admissible inverse branches. Stop concentration interpretation when the signal or model is insufficient; a numerical ratio alone does not establish a physical solution.
\end{enumerate}

Subtraction removes an estimated mean background, not its shot noise or estimation variance. Near zero corrected signal, ratios and their inverse uncertainties can be strongly biased or unbounded. An interior maximum is also possible at $B=0$ for $\rho\simeq0.62$ (Section~\ref{sec:sign_inversion}); a peak-and-fall curve cannot be attributed to background alone or taken as structural recovery. The present cornea analysis remains on its recorded scale: independent background calibration was not supplied and no experimental $\kappa(T)$ is inferred.

\subsection{Conversion-Free Moments and Validation Boundary}
\label{sec:direct_moments}

A spread-to-concentration approximation is unnecessary when a matched distribution of projected axes is available. For independently measured axes $\psi_j$ with stated nonnegative weights $w_j$ of positive total weight, define the exact axial moments and circular variance by
\[
z_n=\frac{\sum_j w_j e^{2in\psi_j}}{\sum_j w_j},
\qquad V_{\rm ax}=1-|z_1|.
\]
For a single axial von Mises population, $z_n=e^{2in\mu}M_n(\kappa)$, so $V_{\rm ax}=1-I_1(\kappa)/I_0(\kappa)$ exactly. These relations require no narrow-distribution approximation, including at small $\kappa$. They describe an orientation distribution; a Fourier moment of the measured excitation-angle intensity is a different quantity until the tensor and detection model are specified. A map of resolved lamellar axes also need not represent subvoxel domain disorder.

Independent physical validation would require matching the structural level, spatial support, weighting and projected geometry of such reference moments to measured SHG harmonics, with independently constrained tensor/background response. Neither fitting $\rho$ and $\kappa$ to the same angular profile nor re-expressing a published three-dimensional cone width supplies that independent reference. Public polarization-stack resources, such as PSHG-TISS \cite{hristu2022pshgtiss}, enable external optical benchmarks, but availability alone does not establish those conditions. No independent matched-disorder validation is supplied in this paper. The direct profile fits in Supplement 1, Section S8, remain a limited compatibility check; they do not validate absolute $\kappa$ or an early thermal-warning window.

\section{Analyzer-Free Angular Measurement and Forward/Reverse Consistency}
\label{sec:experiment}

\subsection{Acquisition Context and Sampling}

The experiment evaluates extraction of analyzer-free angular modulation and consistency between sequential forward and reverse sweeps in prepared porcine cornea. The original and final recording sets are designated corneal acquisitions A and B, respectively; they are analyzed separately. These are technical acquisition datasets, and biological independence is not inferred from their identifiers. Acquisition B supplies the main collagen bundle ROI illustration. Sample rotation changes the nominal tissue/excitation angle in epi collection; it does not calibrate specimen-plane polarization or equal collection weights.

For acquisition B, the operator reported PFA-fixed porcine cornea stored in ethanol, acquired at one unchanged stromal Z at approximately 200--300~$\mu$m. Depth and pixelwise image correspondence were not independently calibrated. Files 1--19 represent $20^\circ$ to $200^\circ$ in $10^\circ$ steps; files 20--37 return from $190^\circ$ to $20^\circ$. The $200^\circ$ image is held out, leaving 36 angular means and 18 matched pairs. Both sweeps use the fixed relative origin $\theta=\theta_{\rm operator}-20^\circ$, without fitted phase alignment. No historical file exclusion transfers to this acquisition.

The bundle ROI crop was fixed at $(320,325)$ in file 1, with offset $(64.5,69.5)$ from the image center, following the existing nominal rotation rule. Its exact implementation was recorded after acquisition and before its angular curve fitting. The ROI uses $112\times112$ pixels and all 225 saved pages per recording, bilinear spatial sampling and float32-to-float64 averaging, retaining signed intensities without additional background correction or per-image normalization. Temporal pages are repeated observations at one angle, not independent complete scans. Full extraction, reference and sensitivity records are given in Supplement 1, Section S9.

The operator confirms that the ROI follows the same tissue across recordings. Small differences in image-plane position, the arrangement and apparent coverage of low-SHG-signal spaces within the crop, and apparent fibre alignment remain. Dark image areas denote low recorded SHG, not verified tissue-free spaces. The whole-ROI mean averages fibre signal together with these low-signal pixels, so changing coverage and alignment may affect mean intensity and angular shape within the same tissue. Crossing fibres further complicate local SHG interpretation \cite{alizadeh2019crossing,alizadeh2023partially}. A single reverse gain need not absorb these image-geometry effects. Their magnitudes and contribution to the observed differences have not been quantified.

\begin{figure}[p]
\centering
{\small Acquisition B: bundle ROI placement across all recordings\par}\smallskip
\includegraphics[width=\textwidth,trim=0 0 0 80,clip]{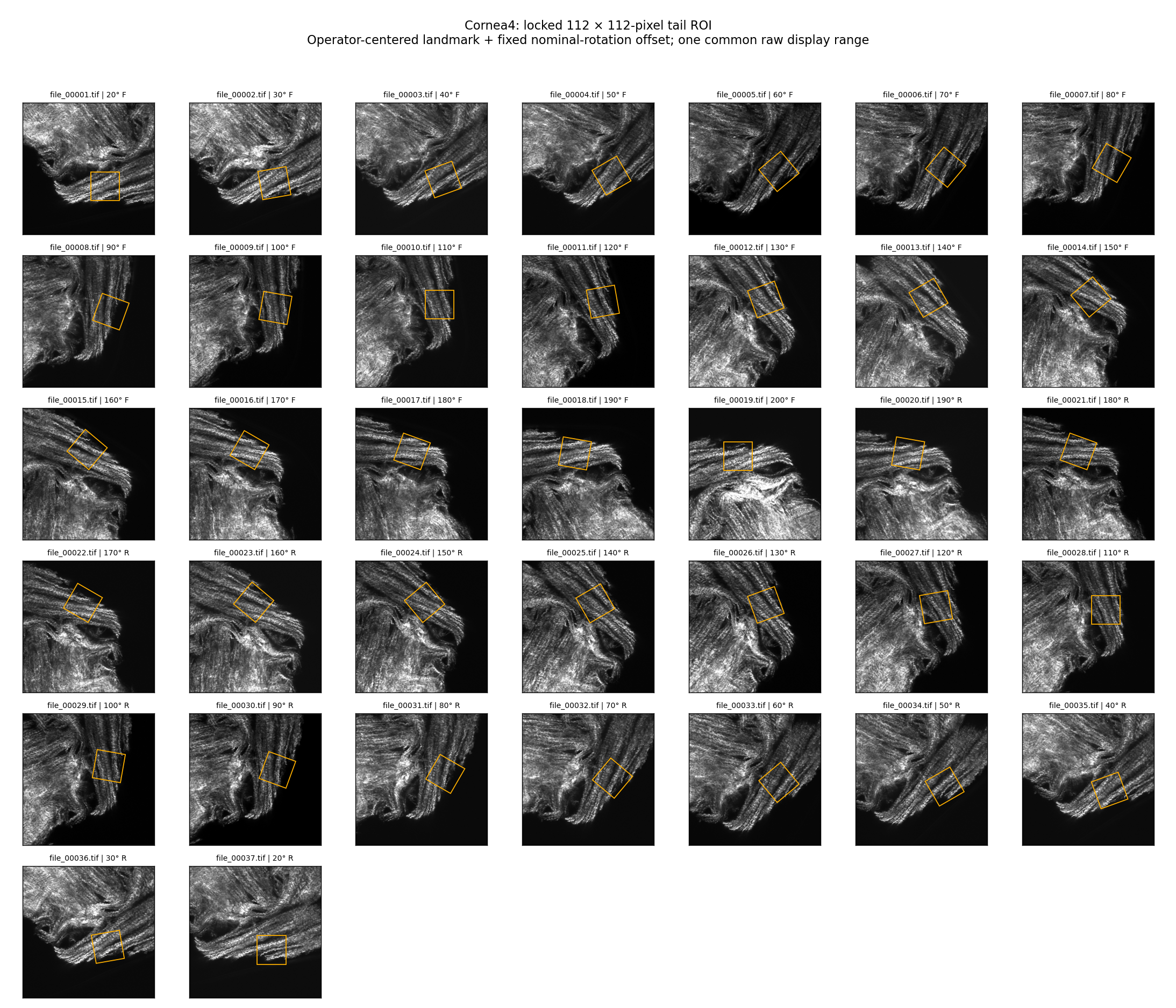}
\caption{Full-field image sampling for acquisition B. Panels follow the operator-declared file sequence: $20^\circ$ to $200^\circ$ forward (F), then $190^\circ$ to $20^\circ$ reverse (R). Each $512\times512$-pixel image averages all 225 saved pages, with one common linear grayscale. Yellow outlines mark the implemented $112\times112$-pixel bundle ROI, using the fixed starting offset from the operator-centered landmark and nominal rotation; no fitted image registration or brightness-based relocation is applied. File 19 at $200^\circ$ is shown for completeness and is held out of curve fitting. The ROI implementation was fixed before its curve fitting, after acquisition. The montage documents implemented sampling of the same operator-identified tissue. Low-signal space coverage and apparent fibre alignment vary slightly; pixelwise correspondence and depth remain uncalibrated. No calibrated physical scale is available.}
\label{fig:roi_overview}
\end{figure}

\begin{figure}[htbp]
\centering
{\small Acquisition B: collagen bundle ROI\par}\smallskip
\includegraphics[width=0.96\textwidth,trim=0 0 0 30,clip]{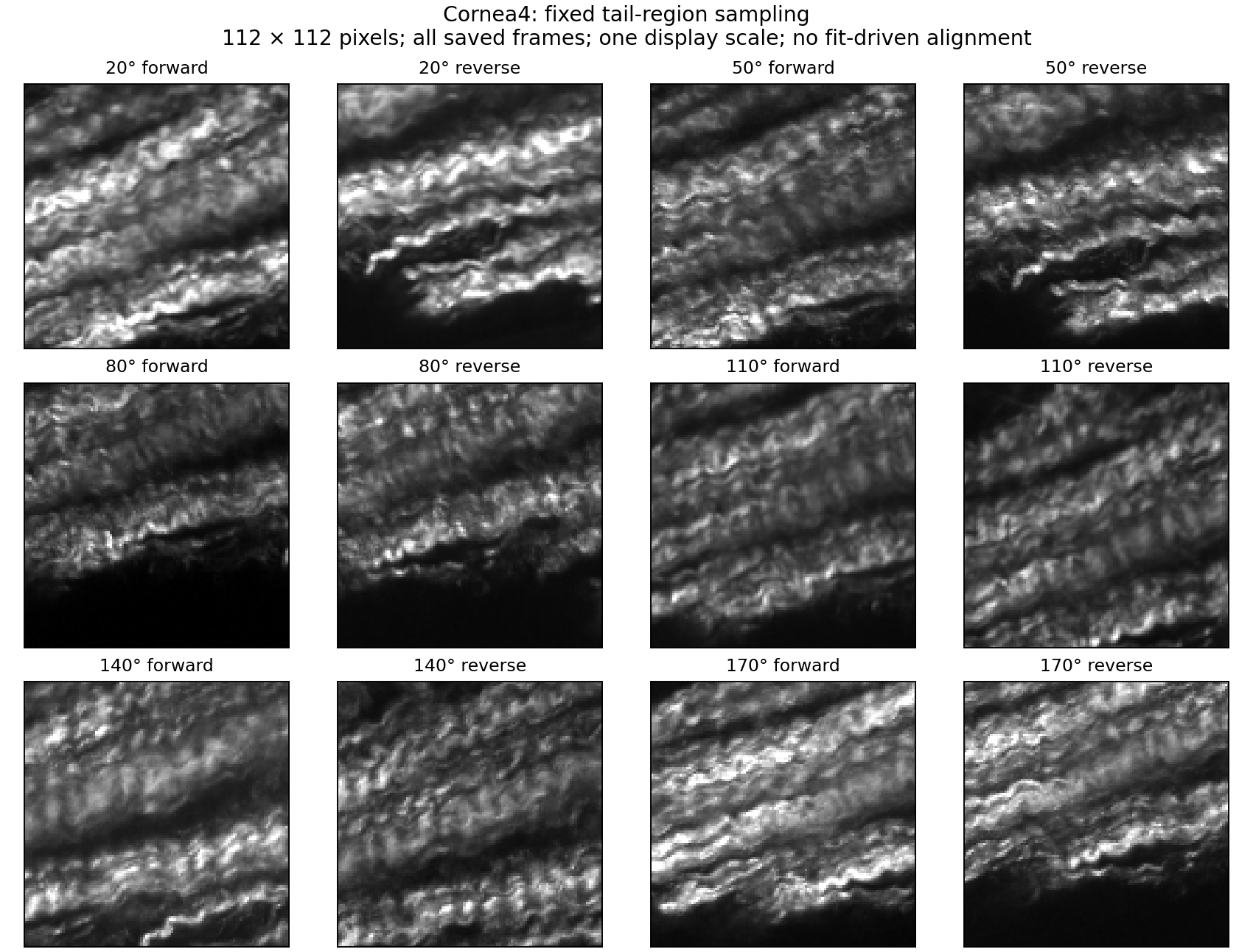}
\caption{Acquisition B, bundle ROI: forward/reverse sampling at six operator-assigned angles. Each $112\times112$-pixel crop averages all 225 saved pages, with one common linear grayscale and no individual brightness normalization. No calibrated physical scale is available. The illustration is post hoc as described in Section~\ref{sec:selection}; the same tissue is operator identified, while pixelwise alignment and depth correspondence are not independently calibrated.}
\label{fig:sampling}
\end{figure}

Figure~\ref{fig:roi_overview} documents the implemented sampling locations in all 37 full-field mean images. Yellow outlines show the $112\times112$-pixel ROI carried through the nominal-rotation rule; the paired crops in Fig.~\ref{fig:sampling} show representative extracted image content.

\subsection{Scale, Normalized Shape and the Primary Test}

The empirical angular function is
\begin{equation}
H(\theta)=a_0+a_2\cos2\theta+b_2\sin2\theta
+a_4\cos4\theta+b_4\sin4\theta.
\label{eq:empirical}
\end{equation}
Its five coefficients are fitted freely; the fit does not impose the physical $\rho$, $\kappa$ or single-axis phase relations of Eq.~\eqref{eq:AF}. Let $\widehat H_F$ and $\widehat H_R$ denote separate forward and reverse fits, with fitted angular means $\widehat a_{0,F}$ and $\widehat a_{0,R}$. The recorded-scale normalized responses are
\[
N_F(\theta)=\frac{\widehat H_F(\theta)}{\widehat a_{0,F}},
\qquad
N_R(\theta)=\frac{\widehat H_R(\theta)}{\widehat a_{0,R}}.
\]
A common multiplicative brightness change cancels in these ratios. Normalized shape still includes modulation depth, angular phase and the relative harmonic contributions; observing modulation in both sweeps alone does not establish equality of these quantities. Because independent background calibration was not supplied, these are normalized recorded responses rather than background-corrected material observables.

The primary restriction is $I_F=H$ and $I_R=gH$, $g>0$, compared with separate five-coefficient curves. Under this restriction, the two entire coefficient vectors share one scale factor; no relative phase is fitted. Same-absolute-curve and relative-affine-offset models are fixed secondary checks. Equal angular weights, restricted HC2 paired-angle wild bootstrap (9,999 draws, seed 2026102903), independent-observation sensitivity, harmonic coefficient screens and the fixed crop/angle checks follow the frozen numerical procedure. The driver completed for acquisition A and the bundle ROI of acquisition B. Pairing uses nominal angle within the same operator-identified tissue. It does not calibrate pixelwise alignment, low-signal space coverage or apparent fibre alignment. The tests and intervals are conditional on the extracted traces and omit unmeasured image-geometry and time-dependent systematics.

This restriction test assesses incompatibility with exact shared shape under its error assumptions. Non-rejection does not establish that all practically important discrepancies are small. No numerical equivalence tolerance was specified; normalized differences, their uncertainty, reference measurements and sensitivity checks are therefore reported alongside the primary $p$ value.

\subsection{Acquisition B: Resolved Modulation and Quantitative Differences}
\label{sec:final_repeat}

Both sweeps resolve the H2 coefficient pair, demonstrating measurable angular modulation in each scan direction. Their normalized H2 amplitudes are 38.37\% forward and 25.36\% reverse; the reverse-minus-forward difference is $-13.00$ percentage points (conditional 95\% interval $-21.85$ to $-4.30$). H4 remains unresolved in both directions. These findings support analyzer-free angular-measurement feasibility while retaining a modulation-depth difference; they do not identify a physical change in fibril orientation or locate a tensor crossover.

The primary shared-scale fit gives $g=0.9632$ (conditional 95\% interval 0.8875--1.0490), paired-bootstrap $p=0.1559$ and independent-observation sensitivity $p=0.0617$. Fit/held-angle RMSE is 383/454 for shared scale and 325/450 for separate shapes. Independent normalization gives RMS/maximal differences of 11.87/21.31 percentage points. The simultaneous 95\% band has a 20.08-point half-width and locally excludes zero near the largest positive difference; its maximum-discrepancy upper bound is 41.39 points. The global restriction test and amplitude/curve diagnostics use different statistics and calibrations, so their mixed evidence is reported together rather than treating one favorable test as a demonstration of agreement.

\begin{figure}[htbp]
\centering
{\small Acquisition B: bundle ROI shared-shape analysis\par}\smallskip
\includegraphics[width=0.98\textwidth,trim=0 0 0 30,clip]{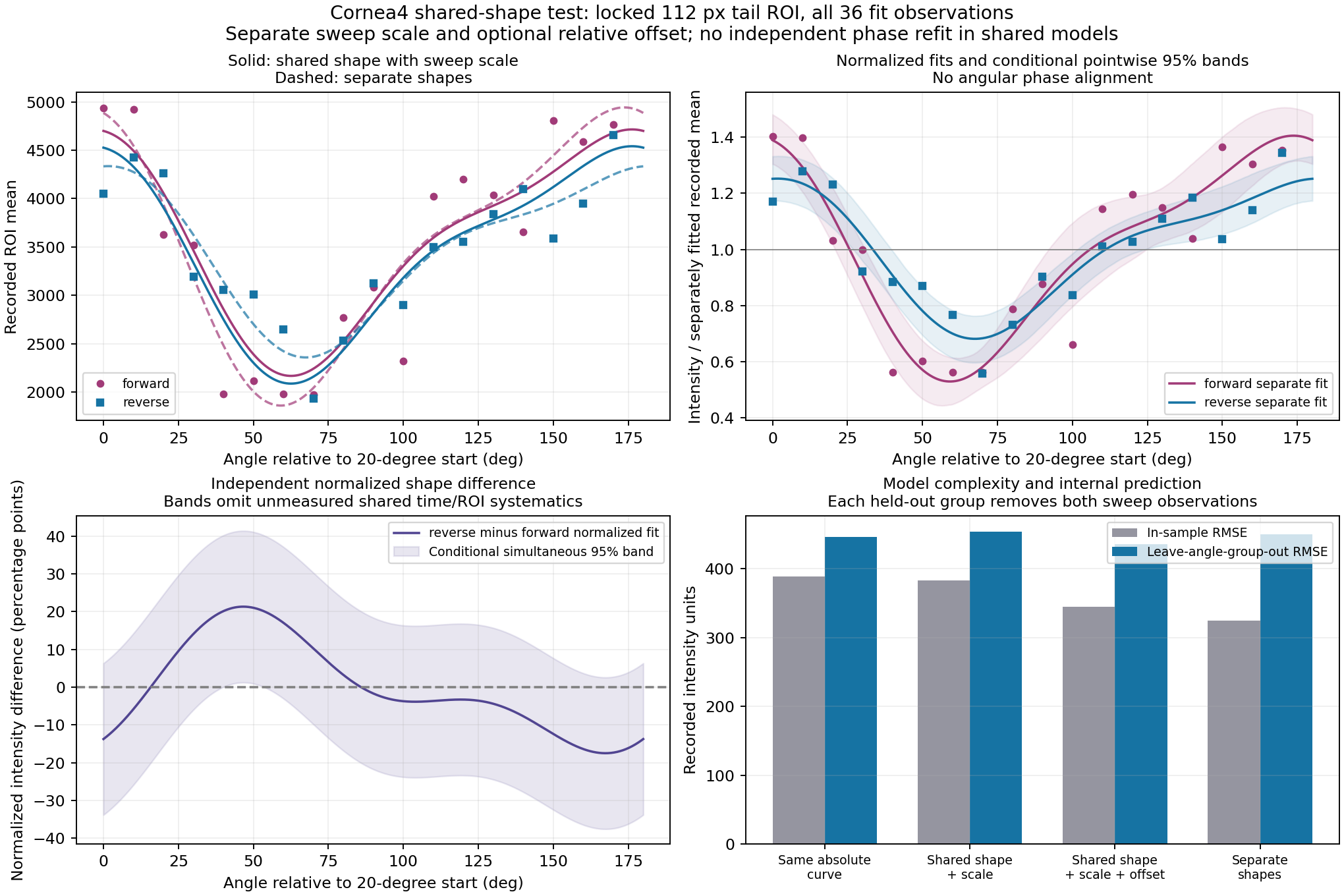}
\caption{Acquisition B, bundle ROI: measured angular means, scale-restricted and separate fits, independently normalized responses, and internal prediction diagnostics. Both sweeps contain resolved H2, while the independently normalized fits retain a modulation-depth difference. Bands are conditional and exclude unmeasured acquisition/geometry systematics. The post hoc illustration does not establish equivalence; selection provenance and all additional outcomes are retained in Supplement 1, Section S9.}
\label{fig:final_repeat}
\end{figure}

The predefined affine secondary model improves on scale-only in its direct paired bootstrap ($p=0.0393$). Its fitted offset is not measured background and does not replace the primary normalization. Across 27 valid crop/translation scenarios, normalized RMS difference spans 10.44--14.30 points and the classical scale-test reference $p$ spans 0.025--0.318. These fixed perturbations are sensitivity checks, not measurements or bounds on the observed recording-specific position, coverage and alignment variations. The held-out $200^\circ$ record is 3.88\% above initial $20^\circ$, while final $20^\circ$ is 17.87\% lower. No middle same-angle reference was acquired. Inconsistent embedded clock ordering is preserved in provenance but does not support reconstruction of elapsed-time drift. Image-position and low-signal-coverage variation, apparent alignment changes, background, temporal variation and intrinsic tissue response are therefore not separately identified. The RMS discrepancies and gain intervals describe the extracted whole-ROI traces and do not isolate a change in intrinsic fibre SHG.

Repeated illumination is a possible exposure-related confound. Collagen SHG can remain stable under scanning conditions \cite{williams2005interpreting}, whereas intense femtosecond irradiation can alter collagen structure and reduce SHG \cite{manickavasagam2014multimodal}. Those exposure conditions do not establish damage in the present fixed-cornea recordings. A constant between-sweep loss is represented by $g$; progressive within-sweep changes can distort normalized angular shape because angle and acquisition order are coupled. Without calibrated specimen-plane power or a dedicated fixed-angle exposure series, exposure-induced tissue alteration cannot be separated from illumination drift, background variation or spatial sampling. Supplement 1, Section S9.C, gives the exposure-related interpretation and its limits.

\subsection{Complete Candidate Outcomes and Presentation Scope}
\label{sec:selection}

Acquisition A provides an additional forward/reverse comparison, with reverse scale 0.8085 (conditional 95\% interval 0.7505--0.8700), primary paired-bootstrap $p=0.1015$ and normalized RMS difference 10.79 percentage points. Only its forward H2 pair is resolved, and both H4 pairs remain unresolved. Its final same-angle reference is 22.13\% lower. Full acquisition A images, model comparisons and uncertainty are reported in Supplement 1, Section S9.A. Table~\ref{tab:roi_selection} summarizes all three candidate analyses, prioritizing effect sizes and conditional uncertainty.

\begin{table}[htbp]
\centering
\caption{All candidate forward/reverse analyses. Reverse gain $g$ has a conditional 95\% interval. RMS is the independently normalized-curve difference in percentage points. Primary paired-bootstrap $p$ values test shared scale versus separate shapes and are nominal, unadjusted for selection. The central ROI is a dependent exploratory partial analysis; its secondary affine verification failed, so its full uncertainty suite is unavailable. These technical comparisons do not establish biological replication or equivalence.}
\label{tab:roi_selection}
{\small
\begin{tabular}{lcrlr}
\hline
Analysis & $g$ (conditional 95\% CI) & RMS & H2 screen & Primary $p$\\
\hline
Acquisition A & 0.8085 (0.7505--0.8700) & 10.79 & Forward only & 0.1015\\
B, bundle ROI & 0.9632 (0.8875--1.0490) & 11.87 & Both sweeps & 0.1559\\
B, central ROI & 1.0614 (0.9773--1.1522) & 14.44 & Neither sweep & 0.0142\\
\hline
\end{tabular}}
\end{table}

The exploratory central ROI combines the operator-reported $-45^\circ$ and $20^\circ$ fibre families and their crossing area. Its normalized RMS/maximal differences are 14.44/23.22 percentage points and its fitted-curve correlation is $-0.663$. The primary paired and independent-observation $p$ values are 0.0142 and 0.0442. All H2/H4 coefficient pairs are unresolved. The primary solver check passes, while the secondary affine verification fails; no simultaneous curve band or full harmonic-difference intervals were obtained. These differing ROI outcomes remain part of the interpretation, with detailed provenance in Supplement 1, Section S9.D. Both B ROIs share recordings.

The bundle illustration was selected post hoc by the largest primary restriction-test $p$. Such outcome-based selection can favor apparent agreement. Its nominal $p$ values and intervals are conditional and unadjusted for selection, tracking, temporal drift or biological variability. The effect sizes and uncertainty therefore carry the empirical interpretation. Non-significance provides no affirmative evidence of repeatability, and no practical equivalence threshold was specified. The complete candidate inventory and selection history remain in Supplement 1, Sections S9.D--S9.E.

\section{Illustrative Thermal Extension}
\label{sec:thermal_extension}

The thermal model illustrates how different assumed trajectories of axial concentration and molecular response can propagate through the harmonic framework. For fixed tensor ratio, detected weighting and background treatment, the angular mean is $S=G\beta^2d_0(\rho)$ for parallel-analyzer detection, or $G\beta^2D_0(\rho)$ for ideal analyzer-free detection, and is independent of pure axial redistribution. This is a model identity under fixed conditions, not an experimental separation of disorder and degradation.

If $\rho$ also changes, even a background-corrected mean at fixed $G$ obeys $S(T)/S(T_0)=[\beta(T)/\beta(T_0)]^2d_0[\rho(T)]/d_0[\rho(T_0)]$ in the parallel geometry, with $D_0$ replacing $d_0$ in the ideal analyzer-free case. Mean intensity alone then cannot isolate the molecular-response trajectory from tensor changes.

The calculation assigns separate Arrhenius trajectories to $\kappa$ and $\beta$. Its threshold separations, Poisson simulations and sensitivity plots are conditional illustrations; they do not identify activation energies, a universal warning window or a molecular-damage temperature. No thermal series is included in the present pilot. The detailed calculation and its assumptions are provided in Supplement 1, Sections S6--S7.

\subsection{Prescribed Thermal History and Transport Limitations}
\label{sec:thermal_transport}

The thermal illustration prescribes a local temperature history $T(t)$; it does not solve heat conduction or infer temperature from the SHG signal. Collagen architecture may influence directional heat transport, but optical orientation anisotropy alone does not determine a thermal-conductivity tensor. Spatial gradients, unequal boundary heating, local optical absorption and, in vivo, perfusion can produce distinct histories in different regions. Applying one imposed ramp to all regions can then bias the kinetics and threshold crossings. No conductivity anisotropy was measured or fitted in this work.

A spatially uniform approximation requires independently justified thermal equilibration. Relevant conditions include a diffusion time $\tau_{\rm diff}\sim L^2/\alpha_{\min}$, using the slowest directional diffusivity and a characteristic length $L$, much shorter than ramp and reaction times, sufficiently uniform boundary conditions, and negligible unresolved local heating. A lumped sample heated from its boundary additionally requires an appropriately small Biot number. These are conditions to verify, not properties established by the present pilot. For a heterogeneous or locally heated specimen, the same local kinetic equations would need independently measured or separately validated $T(\mathbf{x},t)$ rather than the imposed uniform ramp; this paper adds no heat-transfer model.

Temperature errors are amplified by the Arrhenius law. For $k_j(T)=A_j\exp[-E_j/(RT)]$,
\begin{equation}
\delta\log k_j\simeq\frac{E_j}{RT^2}\,\delta T .
\label{eq:temperature_rate_sensitivity}
\end{equation}
At $60^\circ$C, the illustrative activation energies of 100 and 115 kJ/mol give logarithmic sensitivities of approximately 0.108 and 0.125 K$^{-1}$. Averaging a heterogeneous temperature field before evaluating the exponential is generally not equivalent to averaging local rates. Accordingly, the plotted threshold temperatures are conditional examples, not predicted tissue-damage temperatures or validated spatial thermal maps.

\section{Discussion}
\label{sec:discussion}

The analytical contribution is the explicit connection between the local tensor contraction, the detected harmonic orders and the circular moments of a projected axial population. This connection supplies testable interpretation rules: changing the collection geometry changes the local harmonic factors; orientational averaging attenuates those factors through distinct moments; and a zero or turning point can reflect tensor/observable sensitivity rather than absence of orientational structure. The fixed-angle ratio is consequently a secondary, condition-dependent summary of information already present in a full scan.

The numerical and experimental components address different questions. Independent quadrature and inverse checks test the consistency of the stated equations and their implementation. Corneal forward/reverse measurements test extraction of empirical angular modulation and expose acquisition-dependent differences. Neither type of check independently establishes the local tensor ratio, the domain coherence rule or the physical origin of a measured change. This separation allows the analytical result to be assessed without treating a descriptive optical fit as validation of its material parameters.

\subsection{Framework Limitations and Interpretation Rules}
\label{sec:framework_limitations}

The analytical results apply under the following assumptions and boundaries. Each limitation constrains the physical interpretation of an otherwise measurable angular response.

\begin{enumerate}
\item \textbf{Projected, single-population orientation statistics.} The axial von Mises family describes one symmetric population of projected domain axes \cite{mardia2009directional,hillen2017moments}. Crossing populations, out-of-plane structure and a unique three-dimensional distribution are not recovered by this model. Any inferred $\kappa$ refers to this distribution and its assumed weighting; it is not automatically a molecular-order parameter or a concentration transferable from image-derived lamellar axes or three-dimensional cone widths (Section~\ref{sec:direct_moments}).

\item \textbf{Domain-intensity averaging.} Emission is coherent within each domain, but detected domain intensities are added without cross-domain interference. Axial symmetry does not establish this assumption. Relative phase and polarity can matter when domains interfere, so the present average cannot represent all coherent collagen configurations. Independence of the angular mean from pure axial redistribution is conditional on this averaging rule and fixed tensor/collection response.

\item \textbf{Local tensor and optical-response assumptions.} The calculation uses an in-plane tensor with cylindrical and Kleinman symmetry, a common real $\rho\geq0$, and a common response scale. Tensor heterogeneity, symmetry changes, vectorial focusing and propagation effects can alter the coefficients. The analyzer-free formulas additionally assume equal detected weighting of the two in-plane components; removing an output analyzer does not establish that condition in a microscope.

\item \textbf{Conditional parameter identifiability.} A two-angle ratio cannot jointly determine an unknown axis, $\rho$, $\kappa$ and background. Even with fixed $\rho$, saturation and multiple inverse branches can prevent a unique, precise $\kappa$ estimate. Full scans retain more information but do not automatically resolve physical confounding. A fitted effective anisotropy is not independent calibration of the local tensor ratio; concentration inference requires justified tensor/background constraints and an admissible inverse branch (Section~\ref{sec:uncertainty}).

\item \textbf{Geometry-specific criteria and weak harmonics.} Parallel-analyzer and ideal analyzer-free detection retain different harmonic combinations. The endpoint, fixed-angle and precision results for the parallel geometry cannot be transferred directly to analyzer-free measurements. The analyzer-free fourth-harmonic zero at $\rho=1/3$ is conditional; weak order, mixed orientations, optical response or uncertainty can also produce an unresolved fourth harmonic. Its absence does not locate that tensor boundary, and an unresolved coefficient pair has no reliably interpretable phase (Section~\ref{sec:AF}).

\item \textbf{Background, noise and acquisition stability.} Normalization removes a common multiplicative scale, but not additive background or angle-dependent drift. Within the same tracked tissue, image-position variation, changing low-signal space coverage and small apparent alignment changes can affect recorded intensity and angular shape. Existing crop scenarios do not calibrate these effects, and the conditional bootstrap intervals omit their unmeasured contribution. Gain or depth changes can add further confounding. The Fisher bounds and tensor/background bias examples are conditional calculations; they do not establish achievable experimental accuracy or include all calibration, registration and time-dependent errors (Sections~\ref{sec:rho_calibration}--\ref{sec:background_controls}).

\item \textbf{Illustrative thermal interpretation.} The extension prescribes a local temperature history and separate response trajectories; it does not solve anisotropic heat transport or identify those trajectories from data. Changes in $\rho$, gain or background can confound an apparent disorder/degradation separation. Neither a universal temperature cutoff nor an experimentally validated early-warning window follows from these calculations (Sections~\ref{sec:thermal_extension}--\ref{sec:thermal_transport}).
\end{enumerate}

Accordingly, normalized harmonics with uncertainty remain reportable optical observables when physical inversion is unsupported. Assigning their changes uniquely to orientational disorder requires the additional constraints above; where several inverse solutions remain admissible, they should be reported rather than selecting one concentration trajectory.

\subsection{Evidence and Validation Boundary}
\label{sec:validation_boundary}

\textbf{Analytical and numerical support.} The closed-form coefficients follow from the displayed local contractions and axial moments. The independent quadrature, derivative and inverse-root checks in Supplement 1, Section S10, verify their numerical implementation under the stated assumptions. This supports the mathematical relations and their sensitivity examples; it does not establish that one orientation family or coherence rule uniquely describes tissue.

\textbf{Optical observations.} Acquisition B resolves H2 in both directions, while acquisition A supplies a separate technical comparison. Normalized modulation-depth differences, the different central-ROI outcome in Table~\ref{tab:roi_selection}, unresolved H4, ROI sensitivity and same-angle reference changes bound the empirical interpretation. A forward/reverse sweep samples the angular response again and provides a useful consistency check; it is not, by itself, demonstration of an invariant normalized waveform or independent biological replication. Selection provenance and the complete exploratory outcome record are retained in Supplement 1, Section S9.

\textbf{Physical inference.} The digitized SPA comparison in Supplement 1, Section S8 \cite{matteini2012thermal}, shows approximate profile compatibility under a simplified analyzer-free model. Effective fitted parameters and pixel-reading ranges are not independent absolute $\rho$ or $\kappa$ calibration. Such inference requires matched structural-level, tensor/background and optical-response constraints (Section~\ref{sec:direct_moments}). The analyzer-free harmonic zero and thermal examples therefore remain conditional predictions. The current results establish neither the physical cause of the sweep differences nor thermal disorder/degradation separation.

\section{Conclusion}

The axial von Mises framework expresses each SHG harmonic as a tensor-dependent local factor multiplied by a projected axial circular moment. Separate parallel-analyzer and ideal analyzer-free contractions yield distinct harmonic orders and sensitivity conditions. The formulas identify why a fixed-angle ratio can saturate or admit multiple inverse solutions, and why an unresolved harmonic cannot independently determine a tissue susceptibility ratio. The analyzer-free fourth-harmonic zero at $\rho=1/3$ remains a geometry-specific prediction under the stated assumptions.

Prepared-cornea recordings demonstrate empirical analyzer-free angular modulation, with H2 resolved in both sweeps of the bundle ROI. The bundle amplitude difference and the different exploratory central-ROI outcome occur within the same operator-identified tissue, with slight image-position, low-signal-coverage and apparent alignment variation. The measurements assess the implemented whole-ROI procedure; intrinsic-response stability cannot be isolated from these effects. The resulting evidence supports angular-measurement feasibility and the reporting of harmonics with uncertainty, while physical $\rho$, $\kappa$ and thermal-mechanism inference require additional independent constraints. The thermal calculation illustrates the consequences of specified trajectories rather than an experimentally calibrated separation of disorder and degradation.

\begin{backmatter}

\bmsection{Funding} This work was supported by the Research Grants Council of Hong Kong Special Administrative Region, China (project \#: CUHK 14210525 and C5005-23WF) and the Hong Kong Innovation and Technology Fund (Grant No. ITS/392/24).

\bmsection{Acknowledgments} The authors thank Dr. Shiyang Yuan for assistance with operating the SHG microscope.

\bmsection{Author Contributions} Sui Ip Ng and Ken-Tye Yong conceptualized the study. Sui Ip Ng developed the analytical framework and derivations, performed the numerical analysis, experiments and validation, and drafted the manuscript. Yi-Ping Ho provided experimental support and reviewed the manuscript. Ken-Tye Yong supervised the work and reviewed the manuscript.

\bmsection{Disclosures} The authors declare no conflicts of interest.

\bmsection{Data Availability Statement} The author-review handoff includes all 37 original acquisition B microscopy TIFF stacks (225 saved pages each), raw-file SHA-256 hashes, metadata, ROI placement/transforms and per-page ROI traces, alongside the analysis sources and derived tables. Acquisition A is supplied as full-field temporal-mean caches and analysis/provenance records; its original page stacks are not included. The handoff has no public repository accession at this revision. The acquisition B originals and the acquisition A originals may be requested from the authors. The digitized literature comparison is attributed to its published source.

\bmsection{Supplemental Document} Supplement 1 provides the complete algebra, analyzer-free derivation, limits, molecular-pitch relation, background analysis, existing illustrative thermal calculations, digitized literature comparison, technical-pilot analysis details, and numerical tensor/background sensitivity checks.

\end{backmatter}

\bibliography{main}

\end{document}


\maketitle

\section{Orientational Averaging of the Collagen Susceptibility Tensor}
\label{sec:S1}

The local second-order susceptibility tensor for an in-plane collagen fibril with cylindrical symmetry ($C_{\infty v}$) and Kleinman symmetry is expressed in laboratory coordinates as

\begin{equation}
\chi_{ijk}^{\rm local} = \chi_1 u_i u_j u_k + \chi_2(u_i v_j v_k + v_i u_j v_k + v_i v_j u_k),
\label{eq:S1}
\end{equation}

where $\mathbf{u}=(\cos\psi,\sin\psi,0)$ is a representative fibril-axis direction, $\mathbf{v}=(-\sin\psi,\cos\psi,0)$ is the in-plane transverse direction, $\chi_1=\chi_{uuu}$, $\chi_2=\chi_{uvv}$, and $\rho\equiv\chi_2/\chi_1$. The physical angle is $\psi$. Its axial density is
$f_{\rm ax}(\psi\mid\mu,\kappa)=\exp[\kappa\cos2(\psi-\mu)]/[2\pi I_0(\kappa)]$ over a full circle. It satisfies $f_{\rm ax}(\psi+\pi)=f_{\rm ax}(\psi)$, while the local tensor in Eq.~\eqref{eq:S1} changes sign under simultaneous reversal of $\mathbf{u}$ and $\mathbf{v}$. A signed mean tensor therefore vanishes for equal opposite polarities.

We instead assume coherent emission within each domain and negligible detected cross-domain interference. After removing a common intensity scale, the observable is
\begin{equation}
I_\parallel(\phi;\kappa,\rho,\mu)=
\int_0^{2\pi} f_{\rm ax}(\psi\mid\mu,\kappa)
\left|\frac{e_i(\phi)\chi_{ijk}^{\rm local}(\psi)e_j(\phi)e_k(\phi)}
{\chi_1}\right|^2\,d\psi,
\label{eq:S2}
\end{equation}
where $\mathbf{e}(\phi)=(\cos\phi,\sin\phi,0)$ and repeated Cartesian indices are summed. This intensity average is distinct from squaring the mean tensor. The effective number and detected weighting of domains enter an overall scale, not an axially averaged rank-three susceptibility. Intensity addition requires physical justification such as phase averaging or resolved-domain intensity aggregation; axial symmetry alone does not justify neglecting interference.

For $\mu=0$, the relevant axial moments are
\begin{equation}
M_n(\kappa)=\langle\cos2n\psi\rangle_\kappa
=\frac{I_n(\kappa)}{I_0(\kappa)},\qquad
\langle\sin2n\psi\rangle_\kappa=0,\qquad
\langle\cos^{2m+1}\psi\rangle_\kappa=0 .
\label{eq:S3}
\end{equation}
The even physical-angle power moments are
\begin{equation}
\begin{aligned}
\langle\cos^2\psi\rangle_\kappa&=\frac{1+M_1}{2},\\
\langle\cos^4\psi\rangle_\kappa&=\frac{3+4M_1+M_2}{8},\\
\langle\cos^6\psi\rangle_\kappa&=\frac{10+15M_1+6M_2+M_3}{32}.
\end{aligned}
\label{eq:S4}
\end{equation}
Thus the polar moments $A_1$ and $A_3$ from a distribution proportional to $\exp(\kappa\cos\psi)$ cannot be retained. Defining a doubled statistical angle does not change the physical tensor rotation angle.

The local laboratory-frame components, before intensity averaging, are
\begin{align}
\chi_{xxx}^{\rm local}(\psi)&=\chi_1[
\cos^3\psi+3\rho\cos\psi\sin^2\psi],
\label{eq:S5}\\
\chi_{xyy}^{\rm local}(\psi)&=\chi_{yxy}^{\rm local}(\psi)
=\chi_{yyx}^{\rm local}(\psi)\nonumber\\
&=\chi_1[\cos\psi\sin^2\psi+
\rho(\cos^3\psi-2\cos\psi\sin^2\psi)].
\label{eq:S6}
\end{align}
Other local components need not vanish for a rotated domain, but their signed axial averages also vanish. These expressions are used through the local field contraction in Eq.~\eqref{eq:S2}, not as nonzero macroscopic axial-average components.

\section{Algebraic Reduction to Closed-Form Fourier Coefficients}
\label{sec:S2}

For a domain at angle $\psi$, let $\delta=\phi-\psi$ be the physical excitation angle in its frame. The normalized nonlinear polarization components in that frame are
\begin{equation}
P_u^{(2)}\propto\cos^2\delta+\rho\sin^2\delta,\qquad
P_v^{(2)}\propto2\rho\cos\delta\sin\delta.
\label{eq:S7}
\end{equation}
The normalized parallel-analyzer field is
\begin{equation}
g_\rho(\delta)=P_u^{(2)}\cos\delta+P_v^{(2)}\sin\delta
=\cos^3\delta+3\rho\cos\delta\sin^2\delta.
\label{eq:S8}
\end{equation}
Using $\sin^2\delta=1-\cos^2\delta$,
\begin{equation}
g_\rho(\delta)=(1-3\rho)\cos^3\delta+3\rho\cos\delta
\equiv P\cos^3\delta+Q\cos\delta .
\label{eq:S9}
\end{equation}
Here $P,Q$ describe the \emph{local} normalized field and are independent of $\kappa$. Its squared amplitude is
$g_\rho^2=P^2\cos^6\delta+2PQ\cos^4\delta+Q^2\cos^2\delta$.

Using the standard cosine-power identities,
\begin{align}
\cos^2\delta&=\tfrac12(1+\cos2\delta),\label{eq:S10}\\
\cos^4\delta&=\tfrac18(3+4\cos2\delta+\cos4\delta),\label{eq:S11}\\
\cos^6\delta&=\tfrac1{32}(10+15\cos2\delta+6\cos4\delta+\cos6\delta),
\label{eq:S12}
\end{align}
the local intensity has coefficients
$d_0=5P^2/16+3PQ/4+Q^2/2$,
$d_2=15P^2/32+PQ+Q^2/2$,
$d_4=3P^2/16+PQ/4$, and $d_6=P^2/32$.
In particular, the second-harmonic contribution of $2PQ\cos^4\delta$ is $PQ\cos2\delta$, not $PQ\cos2\delta/2$.

Axial averaging uses
$\langle\cos[2n(\phi-\psi)]\rangle
=M_n\cos[2n(\phi-\mu)]$. The ensemble intensity is therefore
$I_\parallel=c_0+c_2\cos2(\phi-\mu)+c_4\cos4(\phi-\mu)+c_6\cos6(\phi-\mu)$, with
\begin{align}
c_0&=d_0=\frac{5P^2}{16}+\frac{3PQ}{4}+\frac{Q^2}{2},\label{eq:S13}\\
c_2&=d_2M_1=
\left(\frac{15P^2}{32}+PQ+\frac{Q^2}{2}\right)M_1,\label{eq:S14}\\
c_4&=d_4M_2=
\left(\frac{3P^2}{16}+\frac{PQ}{4}\right)M_2,\label{eq:S15}\\
c_6&=d_6M_3=\frac{P^2}{32}M_3.\label{eq:S16}
\end{align}
Substituting
\begin{align}
P(\rho)&=1-3\rho,\label{eq:S17}\\
Q(\rho)&=3\rho,\label{eq:S18}
\end{align}
gives explicit functions of $\kappa,\rho$. Equivalently,
$a=3(1+\rho)/4$, $b=(1-3\rho)/4$, and
$g_\rho=a\cos\delta+b\cos3\delta$, so
$d_0=(a^2+b^2)/2$, $d_2=a^2/2+ab$, $d_4=ab$, and $d_6=b^2/2$.
For $\rho=0$, $P=1,Q=0$ and
\begin{equation}
(c_0,c_2,c_4,c_6)=\frac1{32}(10,15M_1,6M_2,M_3),
\end{equation}
recovering the rank-one axial coefficients in the main manuscript.
The polarization mean is $c_0=d_0(\rho)$ and is independent of $\kappa$; multiplying by a common molecular-response intensity scale gives $S\propto\beta^2d_0(\rho)$.

\subsection{Analyzer-Free Detection}
\label{sec:S2AF}

The parallel-analyzer contraction in Eq.~\eqref{eq:S8} must not be used for detection that sums both in-plane polarization components. Assuming equal collection weighting for those components, the normalized local analyzer-free intensity is instead
\begin{equation}
h_\rho(\delta)=
(\cos^2\delta+\rho\sin^2\delta)^2
+4\rho^2\cos^2\delta\sin^2\delta .
\label{eq:SAF_local}
\end{equation}
This expression follows by adding the squared components in Eq.~\eqref{eq:S7}, not by projecting their sum onto the excitation polarization. Its Fourier expansion and axial intensity average give
\begin{equation}
I_{\rm AF}(\phi)=D_0+D_2M_1\cos2(\phi-\mu)
+D_4M_2\cos4(\phi-\mu),
\label{eq:SAF_intensity}
\end{equation}
with
\begin{equation}
D_0=\frac{3+2\rho+7\rho^2}{8},\qquad
D_2=\frac{1-\rho^2}{2},\qquad
D_4=\frac{(1+\rho)(1-3\rho)}{8}.
\label{eq:SAF_coeffs}
\end{equation}
There is no sixth harmonic in this geometry. In the mean-axis frame, the background-subtracted harmonic ratios are $a_2/a_0=D_2M_1/D_0$ and $a_4/a_0=D_4M_2/D_0$, independent of a common intensity scale. The polarization mean is $D_0$, independent of $\kappa$ at fixed $\rho$ and weighting.

For $\kappa>0$ and real nonnegative $\rho$, the fourth harmonic changes sign at $\rho=1/3$, where $D_4=0$. However, $D_2=4/9$ remains nonzero there: eliminating the fourth harmonic does not eliminate orientational sensitivity. At $\rho=1$, the second harmonic vanishes but the fourth remains. Both harmonics vanish in the isotropic limit. These statements apply to a single symmetric axial population and must not be confused with the parallel-analyzer fixed-angle ratio, which remains order-sensitive at $\rho=1/3$.

\begin{figure}[htbp]
\centering
\includegraphics[width=\textwidth]{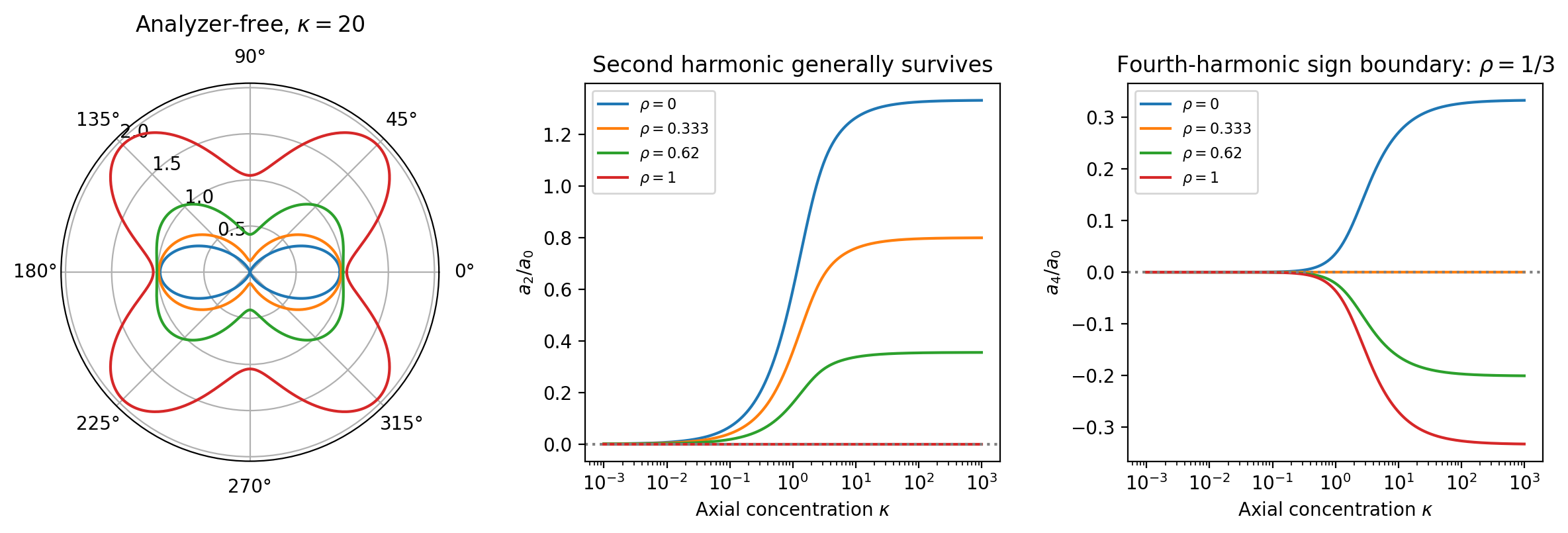}
\caption{Analyzer-free axial-model observables (Figure 11 of the numerical figure set). Left: local-scale ensemble polar patterns at $\kappa=20$ and $\mu=0$ for $\rho=0,1/3,0.62,1$. Middle and right: background-subtracted second- and fourth-harmonic ratios versus axial concentration. The fourth-harmonic sign boundary at $\rho=1/3$ is specific to this geometry and does not imply loss of all orientation information. Equal in-plane collection weighting and the domain-intensity-addition assumption are imposed.}
\label{fig:SAF}
\end{figure}

For the published SPA convention in which $\alpha=0$ is perpendicular to the mean axis, $\phi-\mu=\alpha-\pi/2$, so Eq.~\eqref{eq:SAF_intensity} becomes $D_0-D_2M_1\cos2\alpha+D_4M_2\cos4\alpha$. This is the simplified geometry used for the manuscript's literature-profile comparison. High-numerical-aperture vectorial effects, unequal collection weighting, and out-of-plane structure require additional modeling.

\section{Asymptotic Limits of the Fixed-Angle Ratio}
\label{sec:S3}

With both sampled angles measured relative to the mean axis,
\begin{equation}
C_{\rm raw}(\kappa,\rho)=
\frac{d_0+d_2M_1+d_4M_2+d_6M_3}{d_0-d_4M_2},
\qquad \tilde{C}=\frac{C_{\rm raw}-1}{7}.
\end{equation}

\subsection{\texorpdfstring{Aligned Limit ($\kappa\to\infty$)}{Aligned Limit (kappa to infinity)}}
\label{sec:S3a}

As $\kappa\to\infty$, $M_1,M_2,M_3\to1$ and the intensity reduces to $g_\rho(\phi-\mu)^2$. Since $P+Q=1$ and $P+2Q=1+3\rho$,
\begin{equation}
I_\parallel(\mu)\to1,\qquad
I_\parallel(\mu+\pi/4)\to\frac{(1+3\rho)^2}{8},
\end{equation}
and hence
\begin{equation}
C_{\rm raw}\to\frac8{(1+3\rho)^2},\qquad
\tilde{C}\to\frac{8/(1+3\rho)^2-1}{7}.
\end{equation}
For $\rho=0$ these limits are 8 and 1. For $\rho=0.62$, they are approximately 0.9780 and $-0.00314$. Equality of the aligned and isotropic ratios occurs at $\rho_*=(\sqrt8-1)/3\simeq0.6095$. This endpoint criterion is not a global monotonicity criterion.

\subsection{\texorpdfstring{Isotropic Limit ($\kappa\to0$)}{Isotropic Limit (kappa to 0)}}
\label{sec:S3b}

As $\kappa\to0$, all $M_n\to0$ for $n\geq1$. Thus $I_\parallel(\phi)\to d_0>0$ for every angle, $C_{\rm raw}\to1$, and $\tilde{C}\to0$. This is a finite nonzero intensity, rather than the ratio of two vanishing coherent mean fields.

Using $M_1=\kappa/2+O(\kappa^3)$, $M_2=\kappa^2/8+O(\kappa^4)$, and $M_3=\kappa^3/48+O(\kappa^5)$ gives
\begin{equation}
C_{\rm raw}=1+\frac{d_2}{2d_0}\kappa+O(\kappa^2),
\qquad d_2=\frac{3(1+\rho)(5-3\rho)}{32}.
\end{equation}
For $\rho_*<\rho<5/3$, the initial slope is positive whereas the aligned endpoint is below one, establishing nonmonotonicity and possible inverse-branch ambiguity even with zero background.
At $\rho=1/3$, $P=0,Q=1$ and
\begin{equation}
I_\parallel(\phi)=\tfrac12[1+M_1\cos2(\phi-\mu)],
\qquad C_{\rm raw}=1+M_1 .
\end{equation}
Thus the ratio does not become order-insensitive at $\rho=1/3$.

For a narrow axial distribution,
$\cos2(\psi-\mu)\simeq1-2(\psi-\mu)^2$, so
$\sigma_{\rm 2D}^2\simeq1/(4\kappa)$.
This physical-angle standard deviation is not automatically the spread parameter of a 3D cone model; a projection mapping is needed before comparing concentrations.

\section{Derivation of the \texorpdfstring{$\rho$}{rho}--Pitch Angle Relationship}
\label{sec:S4}

The collagen triple helix comprises three $\alpha$-chains, each containing peptide bonds oriented at pitch angle $\theta$ relative to the helix axis. For a single harmonophore with molecular hyperpolarizability $\beta$ along its axis, oriented at angles $(\theta, \varphi)$ in the fibril frame (where $\varphi$ is the azimuthal angle about the fibril axis), the hyperpolarizability tensor in the fibril frame is obtained by rotating the molecular tensor $\beta_{z'z'z'} = \beta$:

\begin{equation}
\beta_{ijk}^{\rm fibril} = \beta \, R_{iz'} R_{jz'} R_{kz'},
\label{eq:S19}
\end{equation}

where $R$ is the rotation matrix. For a harmonophore at azimuth $\varphi$, the direction cosines are $u_{z'} = \cos\theta$, $u_{x'} = \sin\theta\cos\varphi$, $u_{y'} = \sin\theta\sin\varphi$.

The fibril-frame components of interest are:

\begin{align}
\beta_{zzz}(\theta,\varphi) &= \beta\cos^3\theta, \label{eq:S20}\\
\beta_{zxx}(\theta,\varphi) &= \beta\cos\theta\sin^2\theta\cos^2\varphi, \label{eq:S21}\\
\beta_{zyy}(\theta,\varphi) &= \beta\cos\theta\sin^2\theta\sin^2\varphi. \label{eq:S22}
\end{align}

The macroscopic susceptibility is obtained by summing over all harmonophores. For the three $\alpha$-chains of the triple helix, the azimuthal angles are $\varphi = 0^\circ, 120^\circ, 240^\circ$. Alternatively, assuming a continuous uniform azimuthal distribution (valid for the coiled-coil geometry), we average over $\varphi \in [0, 2\pi)$:

\begin{equation}
\langle \cos^2\varphi \rangle_\varphi = \langle \sin^2\varphi \rangle_\varphi = \frac{1}{2}.
\label{eq:S23}
\end{equation}

The macroscopic susceptibility components are therefore:

\begin{align}
\chi_{zzz} &= N\beta\langle \cos^3\theta \rangle = N\beta\cos^3\theta, \label{eq:S24}\\
\chi_{zxx} &= N\beta\langle \cos\theta\sin^2\theta\cos^2\varphi \rangle_\varphi 
= \frac{N\beta}{2}\cos\theta\sin^2\theta, \label{eq:S25}\\
\chi_{zyy} &= N\beta\langle \cos\theta\sin^2\theta\sin^2\varphi \rangle_\varphi 
= \frac{N\beta}{2}\cos\theta\sin^2\theta = \chi_{zxx}. \label{eq:S26}
\end{align}

The susceptibility ratio is:

\begin{equation}
\rho \equiv \frac{\chi_{zxx}}{\chi_{zzz}} 
= \frac{\frac{1}{2}\cos\theta\sin^2\theta}{\cos^3\theta}
= \frac{1}{2}\frac{\sin^2\theta}{\cos^2\theta}
= \frac{1}{2}\tan^2\theta.
\label{eq:S27}
\end{equation}

The factor $1/2$ arises from the transverse azimuthal second moment, which is the same for three equally weighted chains separated by $120^\circ$ and for a continuous uniform azimuthal distribution. This equality concerns the listed tensor components; a discrete three-chain model does not by itself establish full cylindrical symmetry for every tensor component. The derivation also assumes one rank-one harmonophore type at a common pitch angle, rather than a general mixture of molecular contributions. This coherent local molecular construction is separate from intensity averaging over fibril domains. For native type-I collagen with $\theta \approx 48^\circ$, $\rho = \frac{1}{2}\tan^2(48^\circ) \approx 0.62$.

We note that this derivation assumes an intact triple-helical structure with well-defined cylindrical symmetry. Changes in helix structure, molecular contributions, or azimuthal order can invalidate Eq.~\eqref{eq:S27}; the tensor ratio should therefore be independently calibrated rather than assumed temperature-independent through denaturation.

\section{Background Artifact in the Measured Fixed-Angle Ratio}
\label{sec:S5}

With additive background $B$, a common scale $G$, and molecular response $\beta$, the measured intensities at angles relative to $\mu$ are
\begin{align}
I_{\rm meas}(0^\circ)&=G\beta^2(d_0+d_2M_1+d_4M_2+d_6M_3)+B,\\
I_{\rm meas}(45^\circ)&=G\beta^2(d_0-d_4M_2)+B.
\end{align}
The measured ratio is $C_{\rm raw,meas}=I_{\rm meas}(0^\circ)/I_{\rm meas}(45^\circ)$, with $\tilde{C}_{\rm meas}=(C_{\rm raw,meas}-1)/7$. For nonzero signal,
\begin{equation}
\tilde{C}_{\rm meas}
=\tilde{C}\,
\frac{G\beta^2 I_\parallel(45^\circ)}
{G\beta^2 I_\parallel(45^\circ)+B}.
\end{equation}
Background thus suppresses the magnitude of the fixed-angle metric toward zero. In the isotropic limit, both measured intensities approach $G\beta^2d_0+B$, giving $C_{\rm raw,meas}\to1$ and $\tilde{C}_{\rm meas}\to0$. If signal vanishes at fixed positive background, the measured ratio also approaches one; without signal or background it is undefined.

An interior maximum for $\rho\approx0.62$ can already occur when $B=0$, as shown in Section~\ref{sec:S3}. It cannot be identified as a background artifact solely from its existence. A fixed background also makes the measured ratio depend on $\beta$, so molecular signal loss can mimic reduced anisotropy. Background subtraction or independently calibrated background-aware fitting is needed for the formal separation of orientation from response.

Finally, the background-subtracted polarization mean is
\begin{equation}
S=\frac1\pi\int_0^\pi[I_{\rm meas}(\phi)-B]\,d\phi
=G\beta^2d_0(\rho).
\end{equation}
It is independent of $\kappa$ at fixed $\rho$ and weighting. Consequently, total intensity alone cannot determine the axial orientational trajectory, and the former low-$\kappa$ background limit of $-1/6$ does not apply.

\section{Illustrative Thermal Extension}
\label{sec:thermal}

This section presents a theoretical illustration, rather than an experimentally validated thermal model. For a fixed, known $\rho$ and negligible or subtracted background, the analytical structure provides a formal separation: $\tilde{C}(\kappa;\rho)$ depends on $\kappa$ through the axial moments $M_1,M_2,M_3$, while the background-subtracted polarization mean satisfies
\begin{equation}
S=\frac1\pi\int_0^\pi[I_{\rm meas}(\phi)-B]\,d\phi
=G\beta^2d_0(\rho),\qquad
\frac{S(T)}{S(T_0)}=\frac{\beta(T)^2}{\beta(T_0)^2}.
\label{eq:total_S}
\end{equation}
The normalized relation assumes fixed $G$, $\rho$, and detected domain weighting. Pure axial redistribution does not reduce $S$ in this intensity-averaged model. If $\rho$ changes during helix unwinding, neither the ratio nor the mean uniquely separates disorder and molecular response without further information. A ratio-based $\kappa$ estimate additionally requires an identifiable inverse branch. As an uncalibrated working hypothesis, we use two separate Arrhenius trajectories with linear temperature ramp $T(t)=T_0+\dot Tt$:

\begin{equation}
\frac{d\kappa}{dt} = -A_\kappa e^{-E_\kappa/RT}\kappa, \qquad
\frac{d\beta}{dt} = -A_\beta e^{-E_\beta/RT}(\beta-\beta_{\rm denat}).
\label{eq:arrhenius}
\end{equation}

Illustrative parameters: $\kappa_0=20$, $\beta_0=1$, $\beta_{\rm denat}=0.03$, $A_\kappa=10^{13}$ s$^{-1}$, $E_\kappa=100$ kJ/mol, $A_\beta=10^{14}$ s$^{-1}$, $E_\beta=115$ kJ/mol, $\dot{T} = 1^\circ$C/min. We emphasize $E_\kappa < E_\beta$ is an unverified hypothesis inspired by observations that interfibrillar spread increases before triple-helix unfolding \cite{matteini2012thermal}. Temperature in the Arrhenius exponent is in kelvin. The numerical implementation uses integrated hazards $H_j(T)=\dot T^{-1}\int_{T_0}^{T}A_j e^{-E_j/(RT')}\,dT'$, giving $\kappa=\kappa_0e^{-H_\kappa}$ and $\beta=\beta_{\rm denat}+(\beta_0-\beta_{\rm denat})e^{-H_\beta}$. Concentrations are axial-model values and cannot be carried over from the polar distribution without recalibration. An illustrative threshold separation is defined by first downward crossings of $\tilde{C}_{\rm meas}(T)/\tilde{C}_{\rm meas}(T_0)$ and $S(T)/S(T_0)$ at 0.85 for the rank-one baseline. With $T_0=37^\circ$C and $B=0.02$ in local intensity units, the illustrative crossings are $T_{\rm warn}\simeq56.45^\circ$C and $T_{\rm loss}\simeq65.40^\circ$C, giving $\Delta T\simeq8.95^\circ$C. The ramp extends to $85^\circ$C to include both crossings; absent crossings remain undefined. The inequality $E_\kappa<E_\beta$ alone is insufficient to guarantee a positive window because prefactors, thresholds, and initial conditions also matter.

\begin{figure}[htbp]
\centering
\includegraphics[width=0.85\textwidth]{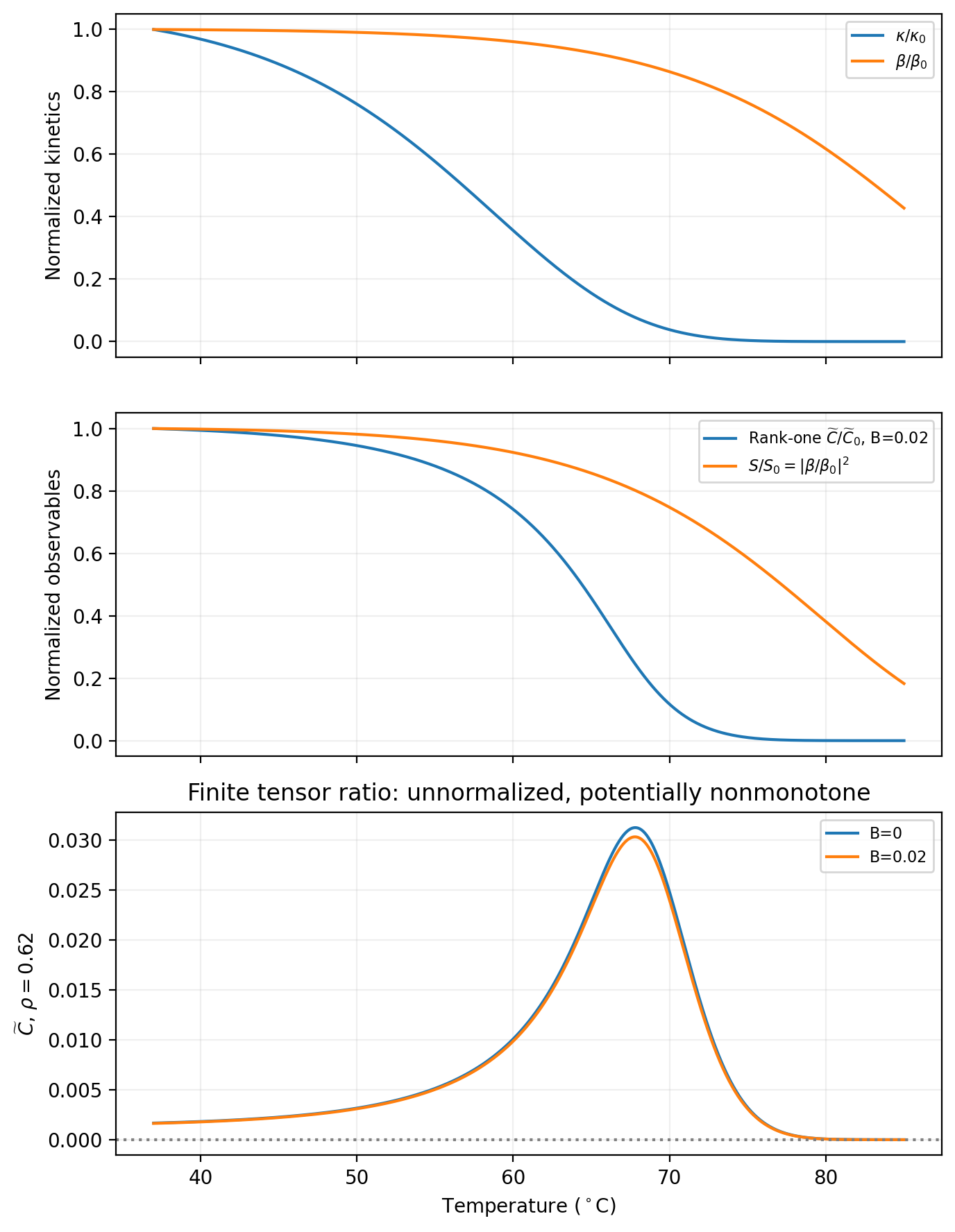}
\caption{Illustrative thermal evolution under the stated kinetic assumptions. (a) Normalised concentration $\kappa(T)/\kappa_0$ and molecular response $\beta(T)/\beta_0$. (b) Initial-value-normalised fixed-angle metric for $\rho=0$ with $B=0.02$ and background-subtracted polarization mean $S/S_0$. Their different trajectories reflect the assumed kinetic hierarchy. (c) Unnormalised fixed-angle metric for $\rho=0.62$, with $B=0$ and $B=0.02$: the response is nonmonotonic even without background, so its direction of change is not a universal indicator of orientational disorder.}
\label{fig:thermal}
\end{figure}

\begin{figure}[htbp]
\centering
\includegraphics[width=0.85\textwidth]{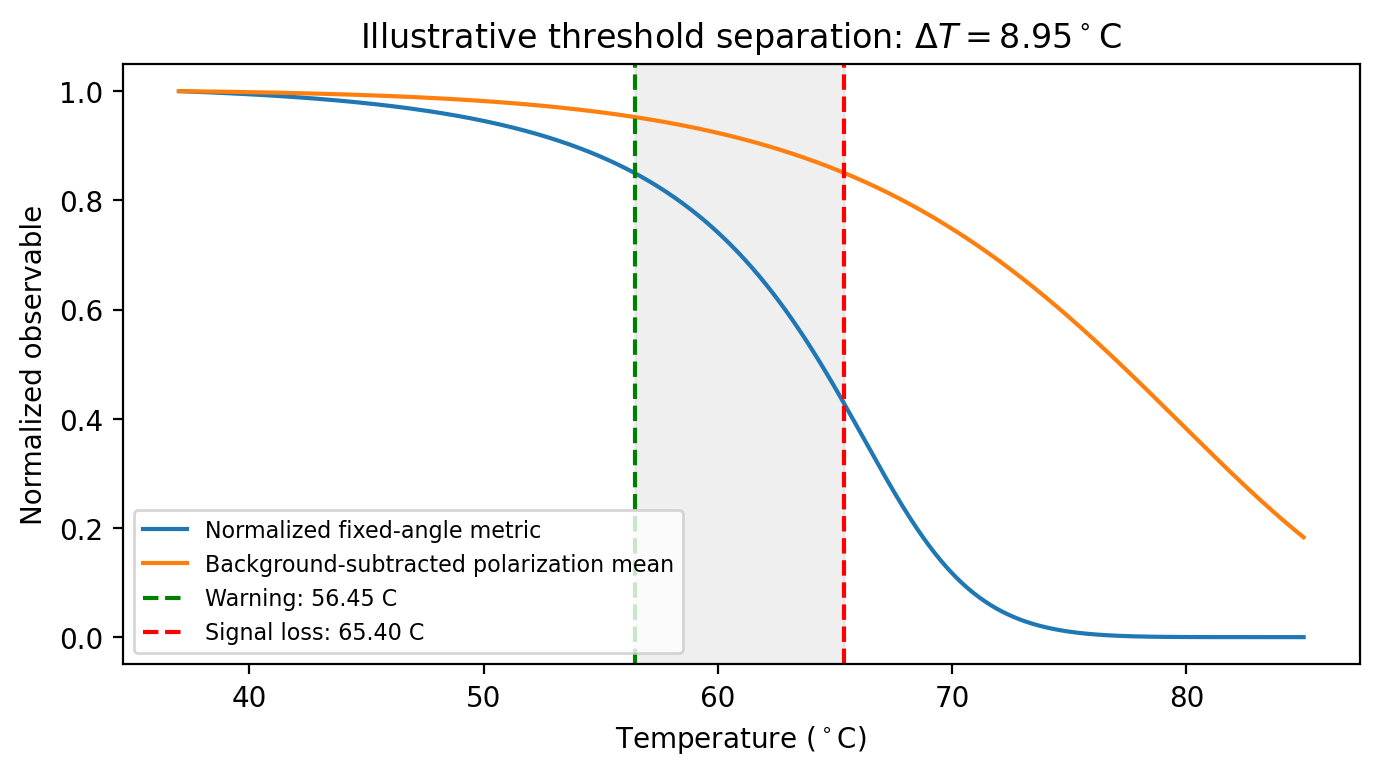}
\caption{Illustrative early-warning threshold separation for the parallel-analyzer rank-one model ($\rho=0$), using the kinetic parameters stated above, a heating rate of $1^\circ$C/min, and background $B=0.02$ in local intensity units. Both the measured fixed-angle metric and the background-subtracted polarization mean are normalized to their initial values; first downward crossings at 0.85 give $T_{\rm warn}\simeq56.45^\circ$C and a signal-loss threshold of $65.40^\circ$C, separated by $\Delta T\simeq8.95^\circ$C. The shaded interval is model-dependent and does not represent an experimentally validated damage-warning window.}
\label{fig:early_warning}
\end{figure}

Figure~\ref{fig:early_warning} illustrates the threshold separation for the rank-one baseline ($\rho=0$). The later crossing is a signal-loss threshold, not an independently established molecular-damage temperature. The window depends on the assumed kinetics and thresholds and is not validated by the published SPA comparison.

\section{Thermal Monte Carlo and Sensitivity Analysis}

The following Monte Carlo calculation measures variability of thermal threshold crossings directly from repeated noisy scans, with background assumed known. It is distinct from the prospective Fisher-bound calculation in the main manuscript. No general 5\% concentration-precision guarantee is assumed. If the Fisher calculation were extended to $\rho=1/3$, free amplitude and background would be degenerate with concentration because the scan contains only a constant and a second harmonic.

A Monte Carlo simulation of 5000 Poisson realizations uses 36-angle scans at 200 temperatures from 37 to $85^\circ$C, initial peak signal counts of $200^2$, a known background of 0.02 local intensity units, and fixed noise-free reference baselines. The measured fixed-angle ratio and background-subtracted scan mean are calculated from the same photon counts, with the baseline kinetics of Section~\ref{sec:thermal}. This yields $T_{\rm warn}=56.09\pm0.44^\circ$C and $\Delta T=9.32\pm0.45^\circ$C (mean $\pm$ s.d.), with both crossings present and positive separation in all 5000 realizations. The central 95\% interval of the separation is approximately $[8.55,10.31]^\circ$C. Signed separations are retained rather than selecting only positive windows. The warning mean lies below the noise-free value of $56.45^\circ$C because noise advances the first-passage crossing; the separation is also biased relative to its noise-free value of $8.95^\circ$C. These numbers are illustrative, conditional on the acquisition model, and do not include kinetic-parameter uncertainty. Systematic errors from polarizer misalignment, laser drift, and sample motion will broaden these distributions in practice.

\begin{figure}[htbp]
\centering
\includegraphics[width=0.85\textwidth]{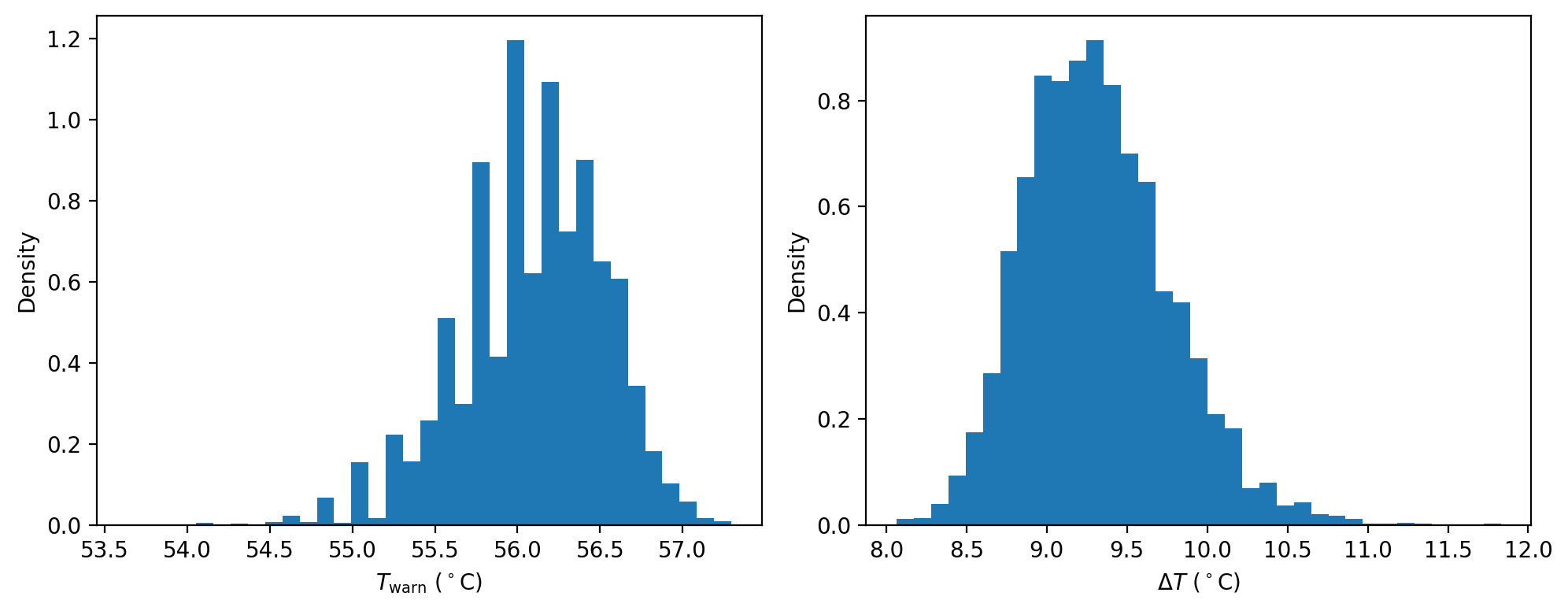}
\caption{Illustrative distributions of $T_{\rm warn}$ and signed $\Delta T$ from 5000 Poisson realizations of 36-angle scans at 200 temperatures. The fixed-angle ratio and background-subtracted scan mean are calculated from the same counts, with known background and fixed noise-free reference baselines. First-crossing noise can bias the threshold temperatures. Kinetic-parameter uncertainty and systematic errors are excluded; the numerical values are conditional rather than predictive.}
\label{fig:temperature_uncertainty}
\end{figure}

\begin{figure}[htbp]
\centering
\includegraphics[width=0.85\textwidth]{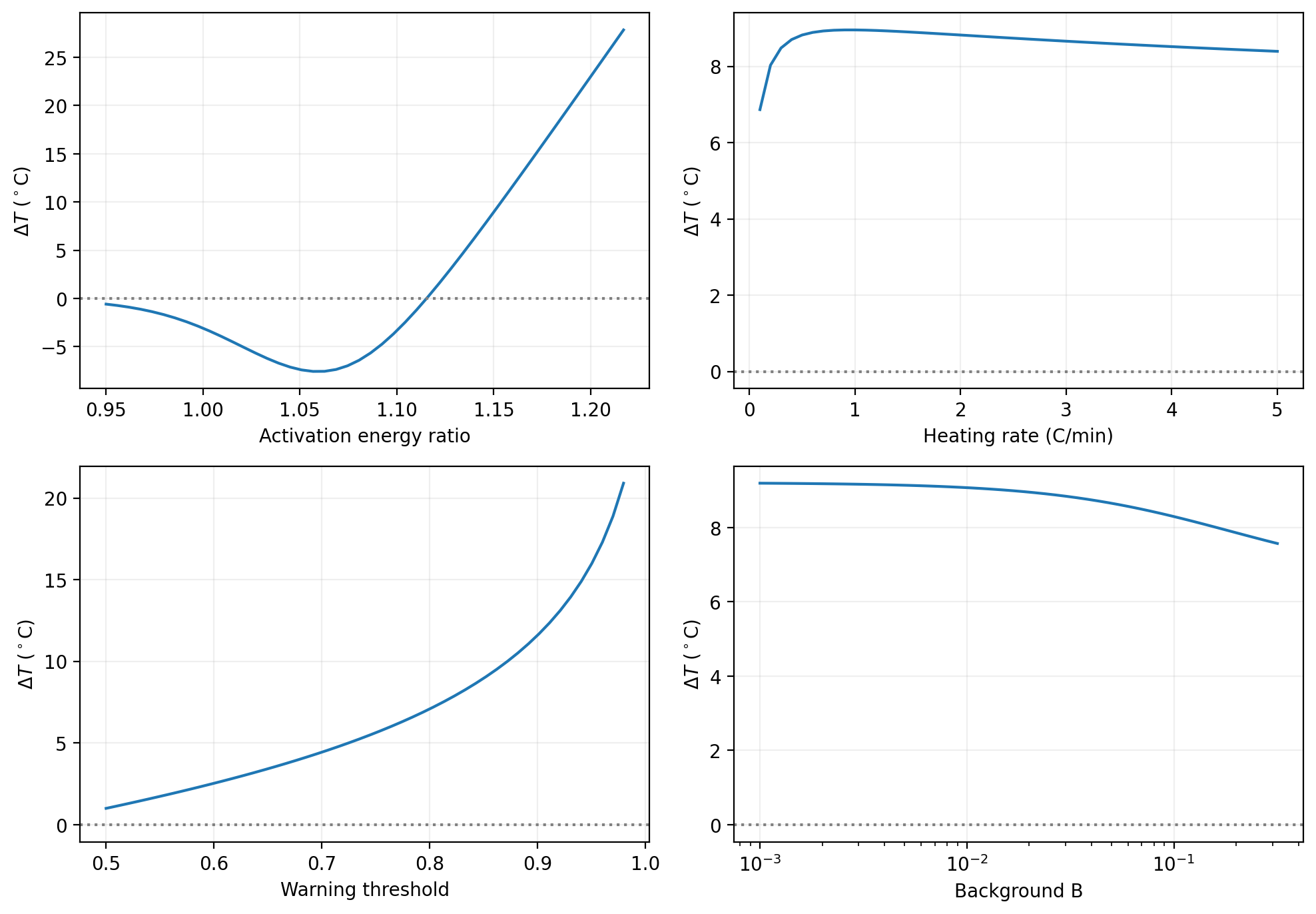}
\caption{Sensitivity of $\Delta T$ to (a) $E_\beta/E_\kappa$, (b) heating rate, (c) fixed-angle ratio threshold $\tilde{C}_{\rm th}$, and (d) background level. Shaded: $\Delta T < 0.5^\circ$C (marginal).}
\label{fig:sensitivity}
\end{figure}

\section{Preliminary Consistency Check Against Published Data}
\label{sec:validation}

\begin{figure}[htbp]
\centering
\includegraphics[width=0.85\textwidth]{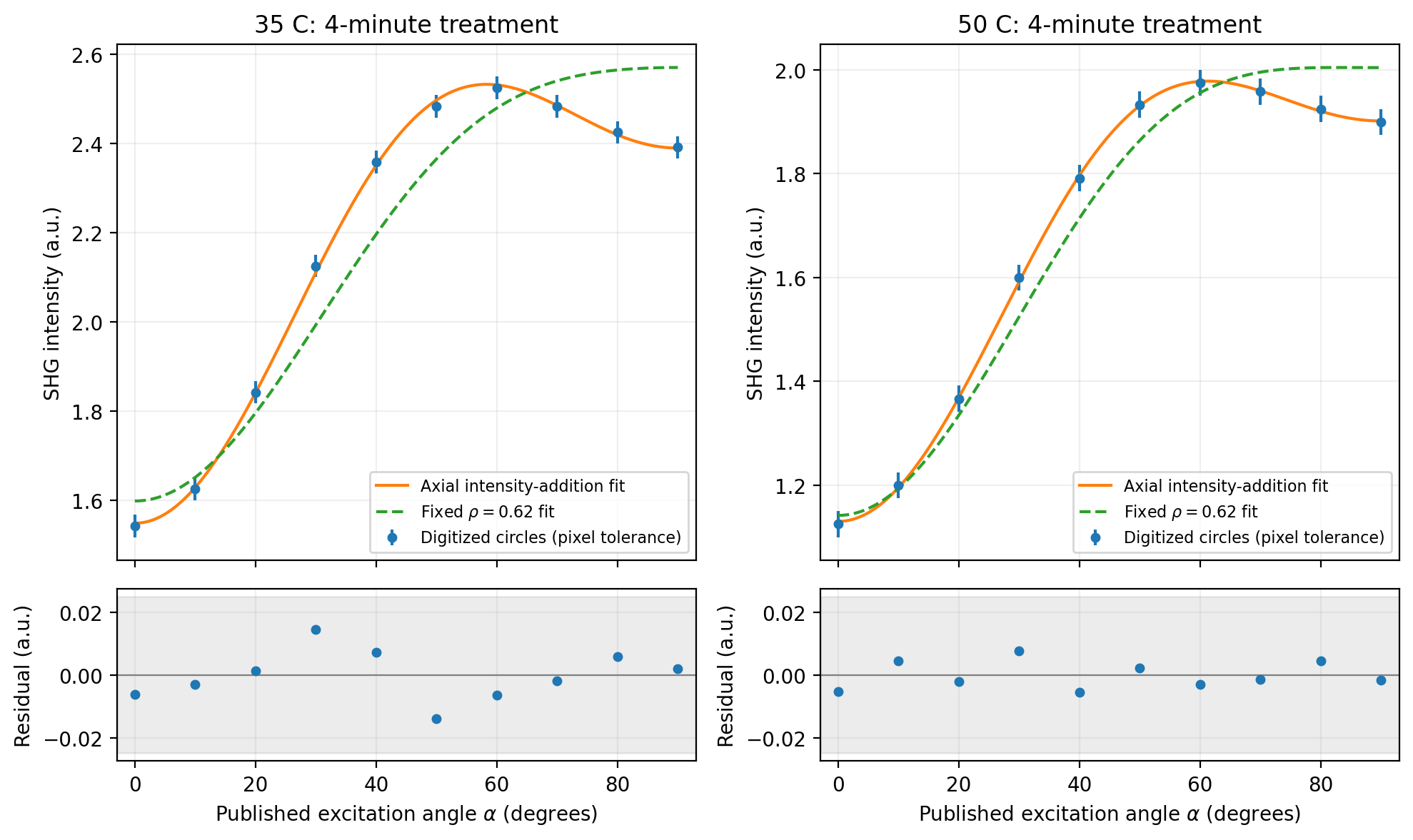}
\caption{Preliminary raster-digitized SPA comparison with Matteini et al.\ \cite{matteini2012thermal}, Fig.\ 4B. Left and right columns show separate 35 and 50$^\circ$C treatments. Upper panels: digitized experimental-circle centers, free-$\rho$ analyzer-free axial fits, and fixed-$\rho=0.62$ fits. Error bars indicate pixel-reading tolerance only. Lower panels: data-minus-free-fit residuals; shaded bands indicate the same tolerance. These approximate fits do not validate the thermal kinetics or a declining axial concentration.}
\label{fig:spa_comparison}
\end{figure}

As a preliminary consistency check, we compare the axial framework with raster-digitized SPA profiles from Matteini et al.\ \cite{matteini2012thermal}. Their corneal samples received separate 4-minute water-bath treatments, rather than a continuous temperature ramp. Figure~\ref{fig:spa_comparison} uses the clearly separated 35 and 50$^\circ$C experimental-circle traces in their Fig.\ 4B; overlapping traces and the logarithmic portion of the intensity axis are excluded. Ten circle centers per trace were read from the public 375-by-237-pixel image, with a stated vertical reading tolerance of $\pm1.5$ pixels, corresponding to $\pm0.025$ intensity units. This tolerance is not an experimental standard error.

The reported forward-detection setup does not describe an output analyzer. We therefore use the analyzer-free derivation and harmonic diagnostics in Section~\ref{sec:S2AF}, rather than the parallel-analyzer contraction. In this geometry, $\rho=1/3$ eliminates only the fourth harmonic, not all orientational sensitivity. We fit the analyzer-free in-plane intensity model
\begin{equation}
I_{\rm AF}(\alpha)=H[D_0-D_2M_1\cos2\alpha+D_4M_2\cos4\alpha],
\end{equation}
where $\alpha=0$ is perpendicular to the mean axis, $H$ is a free nonnegative intensity scale, $D_0=(3+2\rho+7\rho^2)/8$, $D_2=(1-\rho^2)/2$, and $D_4=(1+\rho)(1-3\rho)/8$. Each trace is fitted independently for $H,\kappa,\rho$, with background fixed at zero. The original experiment used 0.95-NA excitation and 1.00-NA collection; vectorial focusing, collection, out-of-plane structure, and the original coherent 3D-cone averaging are not reproduced by this simplified model.

\begin{table}[htbp]
\centering
\caption{Approximate axial fits to digitized experimental circles in Matteini et al.\ \cite{matteini2012thermal}, Fig.\ 4B. Sensitivity ranges are extrema from 300 uniform pixel-tolerance perturbations per trace, not confidence intervals or experimental uncertainties.}
\label{tab:direct_profile_fits}
\begin{tabular}{c|ccc}
\hline
$T$ ($^\circ$C) & Axial $\kappa$ (sensitivity) & $\rho$ & RMS residual (a.u.) \\
\hline
35 & 2.80 (2.61--3.05) & 0.718 & 0.0077 \\
50 & 2.99 (2.74--3.27) & 0.682 & 0.0043 \\
\hline
\end{tabular}
\end{table}

Both free-$\rho$ fits have residuals smaller than the stated pixel-reading tolerance. Fixing $\rho=0.62$ instead gives RMS residuals of 0.110 and 0.061 intensity units, respectively, and fails to reproduce the resolved interior maxima accurately. This supports approximate profile-shape compatibility only when the local tensor ratio is allowed to vary; it does not validate the domain-intensity assumption or determine molecular pitch independently.

These fitted effective parameters are not calibrated absolute tissue values. The fitted axial concentration does not decrease between these two temperatures, and its digitization-sensitivity ranges overlap. Thus this comparison does not establish a decreasing $\kappa(T)$ trajectory or orientational Arrhenius kinetics. In addition, at fixed $\rho$ the polarization mean is independent of $\kappa$, so total intensity supplies no second independent estimate of orientation. The published 3D cone spread must not be converted using $\kappa=2/\sigma^2$; even the narrow-2D relation $\kappa\simeq1/(4\sigma_{\rm 2D}^2)$ requires a justified projection mapping. Raw angular data, calibrated background, and matched optical and thermal conditions are required for stronger validation.

The present calculation fits angular intensities directly and does not use a cone-spread conversion. For a matched, independently observed projected-axis distribution, exact complex moments $z_n=\sum_jw_j e^{2in\psi_j}/\sum_jw_j$ and $V_{\rm ax}=1-|z_1|$ would avoid such a conversion; the single-population relation $V_{\rm ax}=1-I_1(\kappa)/I_0(\kappa)$ is exact. This does not identify subvoxel disorder from a resolved-fiber orientation histogram, nor make a fit of the same SHG profile an independent validation. No second matched dataset or independent concentration reference is analyzed here. The requested validation of absolute disorder or an early-warning window therefore remains outside the evidence supplied by this comparison.

\section{Technical-Pilot Analysis Specification and Scope}

\label{sec:pilot_spec}

The original recording set, designated corneal acquisition A, uses the existing operator-guided $112\times112$-pixel trace and 35 angular means. File 33 (reverse $40^\circ$) was excluded before this shared-shape comparison; no missing measurement is interpolated. Forward $180^\circ$ is held out. Each angular mean contains all 225 saved pages; pages and pixels are not independent angular or biological replicates. Bilinear crop extraction, signed intensities and float32-to-float64 averaging are retained without additional background or specimen-power correction.

The models comprise the same absolute curve (5 mean parameters), a common second/fourth-harmonic curve with positive reverse scale (6), a common curve with positive scale and relative affine offset (7), and separate five-coefficient curves (10). The shared models have no fitted relative angular phase. The affine offset is relative to the forward intercept and cannot be interpreted as an independently measured detector background. All angular means receive equal weight.

The primary comparison uses a restricted HC2-residual wild bootstrap with 9,999 draws and seed 2026102903. One Rademacher sign is shared by observations at the same nominal physical angle. Retain the independent-observation bootstrap as a sensitivity check. The unrestricted separate-shape bootstrap supplies the independently normalized difference curve, its simultaneous 95\% band and amplitude/curve uncertainty on the fixed 721-point $0$--$180^\circ$ grid. These are conditional approximations, not full corrections for unknown time dependence, tracking, spatial error or biological variation. AICc includes the residual-variance parameter, and internal prediction removes the entire nominal-angle group.

The predefined ROI checks comprise concentric squares of 80, 96 and 112 pixels, with center offsets $(0,0)$, $(\pm8,0)$, $(0,\pm8)$, $(\pm16,0)$ and $(0,\pm16)$ raw pixels. Invalid geometries are flagged rather than clipped. Smaller-crop bootstrap checks use 1,999 draws. The existing angle-error sensitivity uses 2,000 independent Uniform($-2^\circ,+2^\circ$) scenarios with coupled crop rotation, plus the nine fixed forward/reverse bias combinations from $\{-2^\circ,0,+2^\circ\}$. These scenarios are assumptions, not calibrated uncertainty distributions or fitted angular alignment. Report matched-angle and all angle-deletion checks without choosing exclusions that improve agreement.

Harmonic resolution is assessed using the existing approximate HC3 joint-quadrature screen, with $W>2F_{2,n-5}(0.95)$. Only the forward H2 pair is resolved in the existing focused comparison. Nonnegative amplitude intervals do not by themselves establish detection near zero; phases for unresolved pairs are not reliable comparisons.

No practical equivalence tolerance was adopted. The assessment therefore remains uncertainty based: non-significance alone does not establish repeatability, and unresolved agreement does not establish intrinsic disagreement. The accompanying procedure froze the numerical choices before the final acquisition below. Reference-only recordings are diagnostic and do not receive extra weight in fitting or define a fitted drift correction. The historical file-33 exclusion does not transfer to new data by filename.

The experimental component is a preliminary analyzer-free feasibility and repeatability assessment. It does not establish absolute $\rho$ or $\kappa$, the exact $\rho=1/3$ crossover, or thermal disorder/degradation separation.

\subsection{Corneal Acquisition A}

The experimental component evaluates whether angular curves can be extracted and whether their normalized shapes are reproducible in the supplied analyzer-free epi-SHG recordings. It is a technical pilot in prepared tissue, not biological replication or parameter calibration. The operator-reported specimen context is PFA-fixed, ethanol-exposed porcine cornea; exact handling, mounting and depth correspondence remain incompletely documented. Sample rotation changes the nominal tissue/excitation angle while collection remains in the epi geometry. A stable unknown angular zero can enter the sine/cosine coefficients, but this does not calibrate specimen-plane polarization or equal collection weighting.

The focused forward/reverse cornea analysis uses one common $112\times112$-pixel square, following the operator-guided anatomical trace with nominal rotation and bilinear spatial sampling. Each angular intensity is the mean of all ROI pixels and all 225 saved temporal pages, retaining signed recorded intensities without additional brightness or background correction. Float32 crop exports are averaged in float64. The saved pages are temporal samples at one angle, not independent complete-scan replicates. Approximate trace placement is a documented limitation rather than independent proof of identical anatomy or depth.

\begin{figure}[htbp]
\centering
\includegraphics[width=0.96\textwidth]{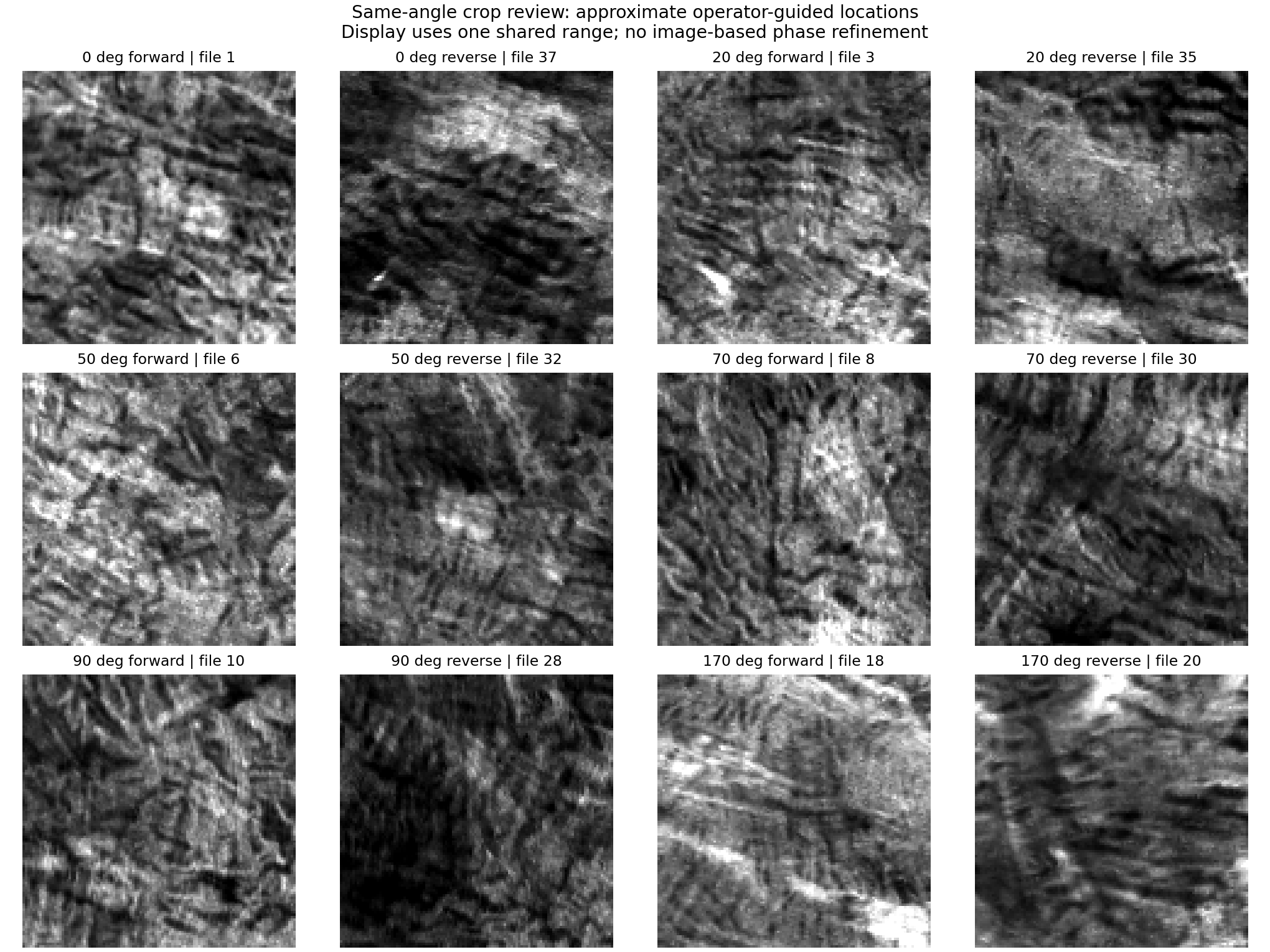}
\caption{Acquisition A: representative sampling images at six nominal angles in the cornea dataset used for the shared-shape analysis. For each angle, forward and reverse $112\times112$-pixel crops are shown side by side after the existing nominal rotation correction. Each image averages all 225 saved pages. A common linear grayscale is used across all panels, without individual brightness normalization. No calibrated physical scale is available. These operator-guided sampling examples do not independently establish anatomical or depth correspondence.}
\label{fig:sampling_cornea2_supp}
\end{figure}

Figure~\ref{fig:sampling_cornea2_supp} shows representative sampled regions from the same forward/reverse cornea dataset, using the recorded operator-guided crop locations and one shared display range.

The primary fit contains 35 angular means: 18 forward angles from $0^\circ$ to $170^\circ$ and 17 reverse angles. The previously excluded file 33 corresponds to reverse $40^\circ$; it is not interpolated. Forward $180^\circ$ is held out as an endpoint check. Equal weights are used because within-file precision does not capture the dominant between-angle systematics. The earlier cornea/muscle feasibility extractions remain separately documented and are not pooled with this focused comparison.

\subsubsection{Scale and Shape Comparison}

The empirical angular function is
\begin{equation}
H(\theta)=a_0+a_2\cos2\theta+b_2\sin2\theta
+a_4\cos4\theta+b_4\sin4\theta.
\label{eq:empirical_cornea2_supp}
\end{equation}
This descriptive fit does not impose the physical single-population analyzer-free parameters derived in Section S2.A. The primary shared-shape model is $I_F(\theta)=H(\theta)$ and $I_R(\theta)=gH(\theta)$, with $g>0$, compared with independent five-coefficient fits in the two directions. A same-absolute-curve model and a shared-shape model with relative affine offset are fixed secondary comparisons. No relative angular phase is fitted to align the sweeps.

The scale-only model estimates $g=0.8085$, with a conditional bootstrap 95\% interval of $0.7505$--$0.8700$. Thus an approximately 19\% overall reverse reduction explains much of the recorded level difference. This fitted scale does not identify laser drift, bleaching, background change or specimen change as its cause.

\begin{table}[htbp]
\centering
\caption{Forward/reverse cornea model comparison. RMSE is in recorded intensity units. The bootstrap is conditional on the stated angular-group error assumptions; it is not an equivalence test.}
\label{tab:pilot_models_cornea2_supp}
\begin{tabular}{lrrr}
\hline
Model & Fit RMSE & Held-angle RMSE & Paired bootstrap $p$\\
\hline
Same absolute curve & 898 & 986 & ---\\
Shared shape + scale & 610 & 733 & 0.1015\\
Shared shape + scale + offset & 585 & 781 & 0.0411\\
Separate shapes & 511 & 722 & ---\\
\hline
\end{tabular}
\end{table}

The primary shared-scale test does not reject its restrictions at 5\%, but this is compatibility rather than proof of repeatability. The result is sensitive to the nuisance model and ROI placement. The secondary affine model has a poorly constrained gain/offset decomposition and does not provide a robust basis for declaring common shape. Its comparison imposes different restrictions and a different fitted null distribution; the two $p$ values are not a ranking of model plausibility. Adding the offset to the scale-only model is not supported by the direct paired-bootstrap comparison ($p=0.4176$).

\subsubsection{Normalized Response and Limits}

The two independent fitted curves are normalized by their own fitted recorded means; the shared model alone cannot demonstrate agreement because it imposes common shape. Their difference is 10.79 percentage points RMS and 20.71 points at its maximum over a full axial period. The conditional simultaneous 95\% band has an 18.05-point half-width; the corresponding upper bound on maximum discrepancy is 38.76 points. These intervals omit unmeasured shared spatial/time systematics. No numerical equivalence tolerance was specified, so the assessment remains uncertainty based.

\begin{figure}[htbp]
\centering
\includegraphics[width=0.96\textwidth]{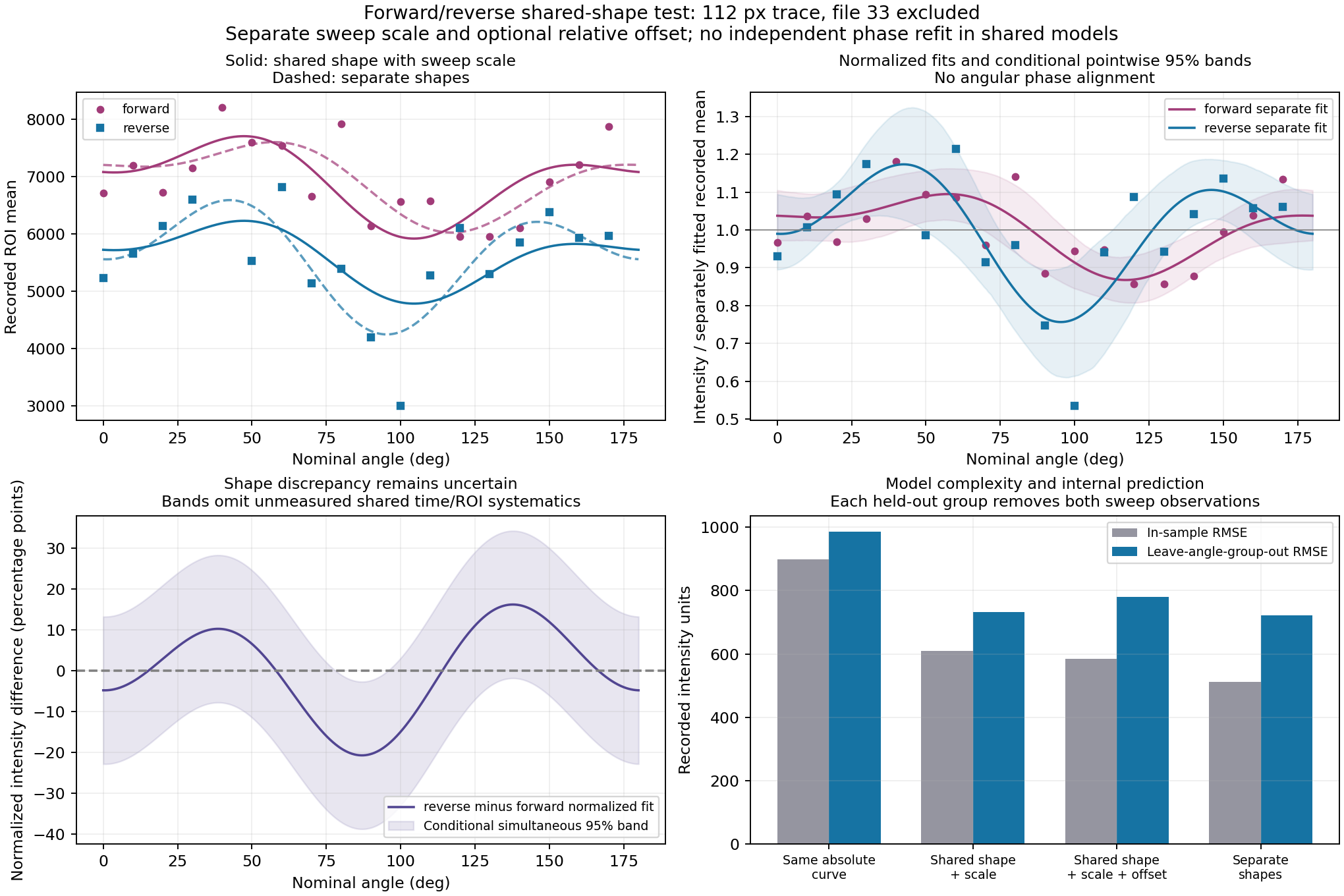}
\caption{Acquisition A: forward/reverse comparison using the fixed 112-pixel trace. A sweep-specific scale explains much of the intensity reduction. Independently normalized fits and their conditional uncertainty retain unresolved shape differences. Held-angle prediction is an internal diagnostic and does not constitute independent experimental validation.}
\label{fig:pilot_cornea2_supp}
\end{figure}

The fourth-harmonic coefficient pair is unresolved in both directions, and reverse H2 is also unresolved under the conditional joint-coefficient screen. Their apparent phase positions cannot establish repeatable phase or physical fibre-axis changes. Across the predefined valid crop/translation scenarios, normalized RMS discrepancy spans approximately 9.28--12.82 percentage points and the shared-scale classical reference $p$ spans 0.008--0.197. These are sensitivity results, not choices for selecting a more favorable ROI.

The forward $180^\circ$ intensity is 5.66\% below initial $0^\circ$, while the final return to $0^\circ$ is 22.13\% lower. No independent middle-of-scan reference was supplied, so temporal drift and angular response are not fully separated in these recordings. The data support preliminary angular-curve extraction and scale-versus-shape assessment. They establish neither reproducible normalized shape nor a significant intrinsic change of tissue angular response. They do not establish absolute $\rho$ or $\kappa$, the $\rho=1/3$ crossover, or thermal disorder/degradation separation.

\subsection{Acquisition B: Input Mapping and Protocol Boundaries}
\label{sec:final_acquisition_spec}

For the final recording set, designated corneal acquisition B, the operator reported PFA-fixed porcine cornea stored in ethanol, imaged at one unchanged Z in the stroma at approximately 200--300~$\mu$m. Fixation/storage durations and independent anatomical depth calibration were not supplied. Files 1--19 represent $20^\circ,30^\circ,\ldots,200^\circ$; files 20--37 represent $190^\circ,180^\circ,\ldots,20^\circ$. The $200^\circ$ record is held out and all other files enter the fit, including file 33. The shared numerical functions use relative angle $\theta=\theta_{\rm operator}-20^\circ$ in both directions. This fixed coordinate origin adds no fitted relative phase or peak alignment. All 18 angle pairs are present; the observed matched design replaces the historical missing-angle case.

The operator confirmed a centered landmark and selected the layered collagen bundle along the lower/right edge in file 1. Before inspecting angular fits, its initial crop center was fixed at $(320,325)$ pixels, with offset $(64.5,69.5)$ from the image center $(255.5,255.5)$. The offset and the $112\times112$ sampling grid follow the frozen nominal handedness $-1$ rotation; no per-angle registration, brightness mask or fit-driven location selection is used. The exact pixel implementation was documented after acquisition, and a pre-acquisition center trace was unavailable. The same tissue is operator identified; pixelwise image alignment and depth correspondence are not independently calibrated. All 27 declared crop/translation scenarios are valid and reported.

Each file contains 225 saved $512\times512$ signed-int16 pages. ScanImage records 900 pre-averaging frames and a logging-average factor of four; the analysis reads all 225 stored pages explicitly. Temporal means use float64, exported crops use float32, and their means use float64, preserving the earlier extraction rule. The author-review handoff includes all 37 original acquisition B TIFF stacks, totaling 8325 saved pages, together with raw SHA-256 hashes, frame metadata, transformed crop coordinates, per-page ROI traces and cache checks. Acquisition A remains available in this handoff through its full-field temporal-mean cache and provenance; its original page stacks were unavailable for inclusion. Public deposition has not been completed. These files enable inspection of sampling and within-recording changes, while the clock discrepancy described below still prevents a reliable between-recording elapsed-time drift reconstruction. Header beam-power control, PMT gains and image geometry are constant across files; channel-offset readouts vary and acquisition offset subtraction is enabled. Instrument readouts are not independent optical-power or tissue-depth calibration. No additional background correction or intensity rescaling is applied.

No middle same-angle reference was recorded. Embedded clock epochs place the two filename blocks in a different order; the operator instructed that their declared filename sequence and directions be used. The original epochs are preserved in provenance, but elapsed-time drift is not reconstructed from them. Initial/final $20^\circ$ observations each receive one angular weight. The held-out $200^\circ$ image tests axial-period endpoint consistency; it is not a separately acquired middle $20^\circ$ reference. These departures limit the claim that the intended acquisition controls were fully documented.

The adapter changes paths, input/design counts, observed matched-angle membership and descriptive metadata only. Ten numerical function ASTs match the frozen source, and all 15 freeze-manifest files remain unchanged. The numerical kernels, models, equal weights, seed, draw counts, angle/crop scenarios and assessment rule are retained. The acquisitions and ROI supports are not pooled and their apparent phases are not aligned. Folder names do not independently establish biological specimen identities.

\paragraph{Same-tissue position, coverage and apparent alignment.}
The operator clarifies that the ROI contains the same tissue across recordings. The observed differences are small changes in image-plane position, the arrangement and apparent coverage of low-SHG-signal spaces inside the crop, and apparent fibre alignment. Low image signal does not establish a tissue-free gap or anatomical replacement. Because the whole-crop mean includes both fibre signal and low-signal pixels, these coverage and alignment variations may affect intensity and angular shape within the same tissue. Crossing-fibre structure can further complicate SHG interpretation \cite{alizadeh2019crossing,alizadeh2023partially}. These effects need not be a constant scale factor. Their size and contribution relative to temporal variation or intrinsic tissue response have not been quantified. Nominal-angle bootstrap pairing does not calibrate pixelwise alignment or low-signal coverage; its intervals omit these unmeasured effects. Exact replay verifies the implemented extraction. The tissue identity is operator reported, and the observed apparent alignment changes do not establish physical fibre reorganisation.

The full-field ROI-location montage is displayed in main Fig.~5, representative bundle ROI crops in Fig.~6, and angular diagnostics in Fig.~7. Full trace records, crops and numerical outputs remain in the reproduction package.

\subsection{Acquisition B: Bundle ROI Results and Stopping Rule}

The primary scale-only fit gives $g=0.9632$ with conditional 95\% interval 0.8875--1.0490 and paired-bootstrap $p=0.1559$ against separate shapes. The independent-observation sensitivity gives $p=0.0617$. Fit/held-angle RMSE values are 389/446 (same absolute curve), 383/454 (shared scale), 345/436 (shared scale plus offset) and 325/450 (separate shapes). The affine secondary comparison with separate shapes gives $p=0.3798$; adding its offset to scale-only gives $p=0.0393$. The affine gain is 0.6382 (conditional interval 0.4314--0.8559), with relative offset 1217.5 recorded units (416.4--1960.5). That fitted nuisance term is not measured background and does not replace the primary normalization.

Independent-fit normalized RMS/maximal differences are 11.87/21.31 percentage points. The conditional simultaneous 95\% half-width is 20.08 points and the maximum-discrepancy upper bound is 41.39 points; zero is locally excluded near the largest positive difference. The global shared-model test and normalized-curve summaries use different statistics and calibrations. Their mixed evidence is retained without creating an equivalence tolerance or choosing a new primary test.

Both H2 pairs pass the fixed coefficient screen. Their normalized amplitudes are 38.37\% (conditional percentile interval 31.63--45.99\%) forward and 25.36\% (19.24--31.94\%) reverse. The difference is $-13.00$ percentage points ($-21.85$ to $-4.30$). Their apparent phase difference is $6.52^\circ$ with conditional draw range $-2.76^\circ$ to $15.96^\circ$, not a calibrated fiber-axis rotation. H4 amplitudes are 12.33\% and 8.08\%, but both coefficient pairs remain unresolved. Positive amplitude-percentile intervals do not establish the sign or phase of a weak harmonic.

Across all 27 geometry-valid ROI scenarios, normalized RMS difference spans 10.44--14.30 points and the classical scale-test reference $p$ spans 0.025--0.318. The independent coupled angle-error cases give central 95\% scale-gain range 0.9568--0.9692 and classical reference-$p$ range 0.0379--0.1059; these are sensitivity ranges, not confidence limits. These fixed crop sizes and $\pm8$/$\pm16$-pixel translations do not quantify the observed recording-specific position, low-signal-coverage and apparent alignment variations. No scenario is selected to improve agreement. The original multistart nonlinear and matched-design SVD checks agree in SSE to better than $1.2\times10^{-7}$ recorded squared units.

Under the operator-declared sequence, initial $20^\circ$, held-out $200^\circ$ and final $20^\circ$ means are 4934.21, 5125.63 and 4052.28 recorded units. The held-out axial endpoint differs by $+3.88$\%, while the two nominal $20^\circ$ records differ by $-17.87$\%. These checks cannot isolate drift, background, sampling or intrinsic tissue change. The final acquisition strengthens H2 measurement feasibility but does not establish reproducible normalized response. The planned experimental expansion is closed; the observed response and limitations are reported without a rescue scan series, known-offset experiment or further theoretical expansion.

\paragraph{Exposure-related alternatives.}
Collagen SHG can be photostable under scanning conditions, although prolonged stationary illumination can produce optical changes \cite{williams2005interpreting}. Under controlled femtosecond laser-ablation conditions, Manickavasagam et al.\ observed collagen structural alteration and SHG loss, with effects depending on irradiation conditions \cite{manickavasagam2014multimodal}. These results motivate cumulative illumination as a possible confound; their specimens and exposure conditions do not demonstrate photodamage or a damage threshold in the present PFA-fixed, ethanol-stored cornea. SHG loss alone does not identify a fluorescence-style photobleaching mechanism.

A constant between-sweep intensity factor is already included in $g$, whereas progressive changes within a sweep can affect normalized shape because acquisition order and angle are coupled. The recorded beam-power control values are not calibrated specimen-plane power, and no dedicated fixed-angle exposure series or reliable elapsed-time trajectory was supplied. The bundle ROI's endpoint signal decreases by 17.87\% at its $20^\circ$ reference, while the central ROI from the same recordings increases by 7.78\% when both central families return to their starting references (Section~\ref{sec:center_exploratory}). Local fibre organisation and polarization response can contribute to ROI-dependent signals \cite{stoller2002polarization,williams2005interpreting}; neither ROI supplies an independent illumination-stability control. The opposite endpoint changes cannot isolate illumination drift or exclude cumulative-exposure effects. Exposure-induced tissue alteration, background variation, focus and spatial-sampling changes remain unresolved contributors.

\subsection{Acquisition B: Central ROI Exploratory Partial Analysis}
\label{sec:center_exploratory}

The operator identifies two fibre families in the central ROI. The diagonal family marked blue starts at $-45^\circ$ and the transverse family marked green starts at $20^\circ$. The green region includes their bright crossing area. Their axial separation is $65^\circ$. Both follow the same rotation increments and return to their starting orientations. The left-family angle follows $-45^\circ\to135^\circ\to-45^\circ$, while the right family and the bundle/tail follow $20^\circ\to200^\circ\to20^\circ$. On a $0$--$180^\circ$ axial display, $-45^\circ$ is represented by $135^\circ$. The assigned directions are operator reported. Analysis uses rotation from the first recording, and historical fields named for $20^\circ$/$200^\circ$ use the tail-coordinate convention. The whole central ROI combines both families.

After the bundle ROI and an initial fixed-image-center implementation had been analyzed, the operator supplied green-circle annotations identifying the intended central anatomy. The fixed-image-center implementation did not target those marks and is retained as superseded for anatomical reasons, not as a fourth ranked candidate. The corrected centers were visually transcribed from the displayed 1696-by-1454-pixel montage into the original 512-by-512 images, using each panel's approximately 192-pixel image extent. This is approximate manual anatomical tracing, not image-registration calibration. Files 6, 7, 13, 14 and 21 had no green mark; only their center coordinates were linearly interpolated between adjacent marked centers within the same sweep. All 37 angular intensities were extracted from measured images; none was interpolated. The per-file trace and provenance were fixed before the corrected fit and are included in the ROI record. The crop side, nominal rotation, signed-intensity averaging and 27 sensitivity scenarios were unchanged.

\begin{figure}[htbp]
\centering
{\small Acquisition B: central ROI (exploratory)\par}\smallskip
\includegraphics[width=0.96\textwidth,trim=0 0 0 30,clip]{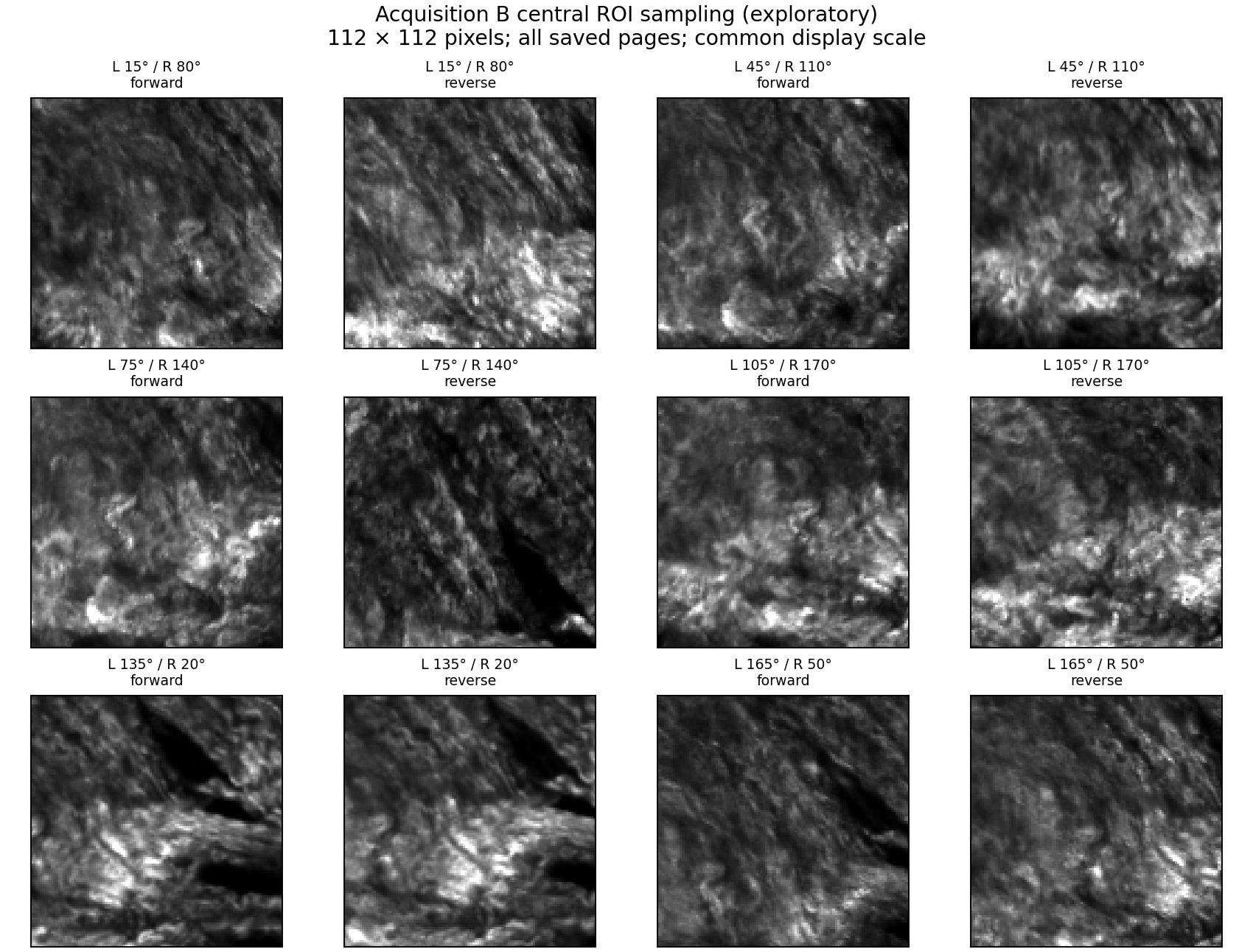}
\caption{Central ROI of acquisition B, an exploratory ROI correction after prior analyses. Panel labels L and R denote the operator-reported fibre families starting at $-45^\circ$ and $20^\circ$, respectively. The annotated diagonal family corresponds to L and the transverse/crossing region to R. Panels are ordered by the L-family axial angle over $0$--$180^\circ$. Forward/reverse crops use the same $112\times112$ support, all saved pages and common grayscale. The per-file manual trace approximates the operator's green circles; five unmarked centers were interpolated geometrically. This is not a new acquisition, an independent specimen or a confirmatory ROI.}
\label{fig:center_sampling}
\end{figure}

The full frozen driver halted at its existing affine-model solver-verification guard. The secondary positive-gain profile reached the fixed upper gain boundary of 100; its SSE differed from the independent nonlinear solver by 38357.53 recorded squared units, exceeding the original 0.01 tolerance. The failed log and partial outputs are preserved. No boundary extension, offset correction, location search or verification-guard removal was used to obtain a successful full run.

A separate failure-aware report calls the original frozen numerical functions without editing them. The primary scale-only fit passes the original three-start solver guard. Its paired and independent wild-bootstrap calls use the original seed and the same first two call positions as the full driver. The primary estimate is $g=1.0614$ (conditional 95\% interval 0.9773--1.1522), with paired $p=0.0142$ and independent-observation sensitivity $p=0.0442$. These are the original primary comparison, not a replacement affine model or a new optimized analysis. The affine p value and regular interval are not reported. The full simultaneous-band/harmonic-interval/angle-error suite was not completed, and this partial result must not be described as full frozen-pipeline success.

Independent-fit normalized RMS/maximal differences are 14.44/23.22 percentage points; the centered coefficient correlation is $-0.663$. H2 amplitudes are 9.77\% forward and 10.01\% reverse, and H4 amplitudes are 5.87\% and 4.53\%. All four coefficient pairs are unresolved by the original joint screen. Opposite fitted H2 trends are point estimates, not evidence of a resolved physical phase reversal. Fit/held-angle RMSE values are approximately 702/851 (same absolute), 683/852 (shared scale) and 563/779 (separate shapes); machine-readable results retain the exact values.

\begin{figure}[htbp]
\centering
{\small Acquisition B: central ROI primary analysis (exploratory)\par}\smallskip
\includegraphics[width=\textwidth,trim=0 0 0 30,clip]{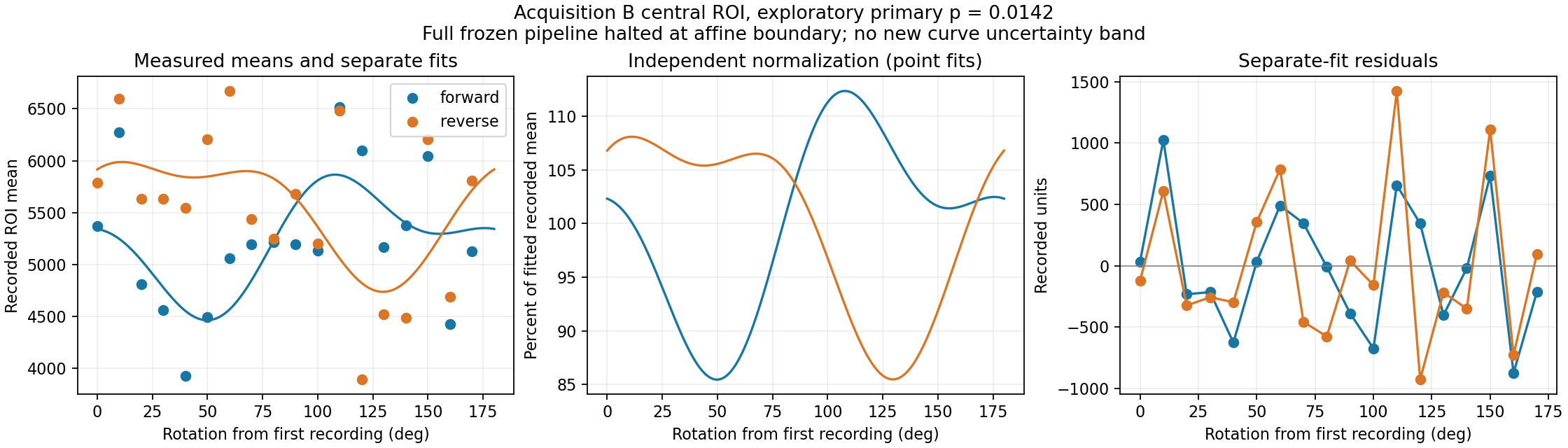}
\caption{Corrected central ROI: measured means, independent normalized point fits and residuals. Horizontal axes show rotation from the first recording, where the left and right families start at $-45^\circ$ and $20^\circ$, respectively. The primary shared-scale restriction has nominal paired $p=0.0142$; the full driver halted on the secondary affine-model verification. No new simultaneous confidence band is constructed. Weak unresolved harmonics, manual correspondence and post hoc ROI definition limit physical interpretation.}
\label{fig:center_primary}
\end{figure}

All 27 central crop/translation scenarios are geometrically valid. Their normalized RMS point discrepancy spans 10.49--19.40 percentage points and classical reference $p$ spans 0.0024--0.2126. These fixed perturbations assess the chosen extraction scenarios; they do not calibrate the observed position, low-signal-space coverage or apparent alignment changes within the same tissue. They are not candidates for choosing a better primary result. The final central recording, with both families returned to their initial references, is 7.78\% above its initial value, differing from the bundle ROI's reduction at its own $20^\circ$ reference within the same recording set. These local-angle references must be distinguished; their ROI dependence prevents interpreting the endpoint changes as a single independently measured drift correction.

\subsubsection{Axial Reordering and Crossing-Fibre Exploration}

A subsequent operator-requested exploration displays the saved observations in ascending axial angle over $0$--$180^\circ$, using both reported fibre-family references for the central trace. The central whole-crop forward/reverse normalized RMS discrepancy is 14.44 percentage points with curve correlation $-0.663$. Approximate polygon masks were manually transcribed from the operator's annotated reverse image (file 23) into the co-rotating 112-pixel crop. The upper blue mask contains 9520 pixels and the lower green mask 2202 pixels. The green mask combines the $20^\circ$ family with the crossing area. Its mean is brighter than the blue-region mean in 35 of all 37 recordings. Their first-to-last changes are $+17.16$\% and $+4.64$\%, respectively. Moving the relevant mask boundary by $\pm4$ pixels gives green-region changes of $+13.09$--$+21.68$\% and blue-region changes of $+3.78$--$+4.88$\%. These post hoc geometric scenarios depend on the manual trace and do not propagate unmeasured recording-specific position, low-signal-coverage or apparent alignment variations within the same tissue. Descriptive H2/H4 regional fits provide exploratory point estimates.

Crossing fibre families can alter polarization-derived collagen parameters \cite{alizadeh2019crossing,alizadeh2023partially}. Under the stated intensity-addition model with equal weights and a common local response, a $65^\circ$ separation gives harmonic retention factors $|\cos65^\circ|=0.423$ for H2 and $|\cos130^\circ|=0.643$ for H4. These conditional examples motivate a mixed-population interpretation of weak central modulation. Family weights, coherent effects, out-of-plane structure, spatial correspondence and temporal changes remain unresolved. The sorted images, curves, mapping, annotation masks and regional results accompany the requested exploratory analysis in the handoff package.

\subsection{Post Hoc Presentation Choice and Dependence}

Main Table 2 displays all three candidate analyses with gain intervals, normalized RMS discrepancies and nominal primary tests. The bundle ROI of acquisition B remains the main image illustration. The three candidate analyses were acquisition A ($p=0.1015$), the bundle ROI of acquisition B ($p=0.1559$) and its exploratory central ROI ($p=0.0142$). The largest primary paired-bootstrap $p$ determined the post hoc illustration choice. Outcome-based selection can favor apparent agreement. All values and conditional intervals are unadjusted for that selection. Effect sizes, uncertainty and the differing central-ROI outcome carry the interpretation; non-significance supplies no affirmative repeatability evidence. No practical equivalence threshold was specified.

The complete central-ROI result remains in Section S9.D because it formed part of this comparison. Its primary comparison is supported by the unchanged routine, while the failed secondary/full-pipeline verification remains unresolved; supplementary placement does not change that partial-analysis boundary. The bundle and central ROIs share raw recordings and cannot count as independent repeat specimens. The earlier forward-only pilot has no reverse sweep and is not an additional candidate. Full acquisition A and bundle-ROI results, including their uncertainty and acquisition limits, remain in Sections S9.A--S9.C.

\section{Numerical Sensitivity to Tensor and Background Calibration}
\label{sec:S10}

\subsection{Tensor Misspecification and Inverse Sets}
\label{sec:S10rho}

These additional deterministic checks evaluate the existing parallel-analyzer equations; they do not fit the cornea data, change the frozen experimental procedure or validate thermal kinetics. Nominal ratios are $\rho_0=0.2,1/3,0.62$, true concentrations are $0.5,2,5,10,20$, and true ratios are $\rho_0(1+\delta)$ with $\delta=-0.1,0,0.1$. Observed noise-free metrics are inverted with $\rho$ fixed at $\rho_0$. Relative perturbation of $\rho_0=0$ would be identically zero and is not used to assess tensor uncertainty. These chosen values are illustrative scenarios, not measured tissue ranges or confidence intervals.

\begin{figure}[htbp]
\centering
\includegraphics[width=\textwidth]{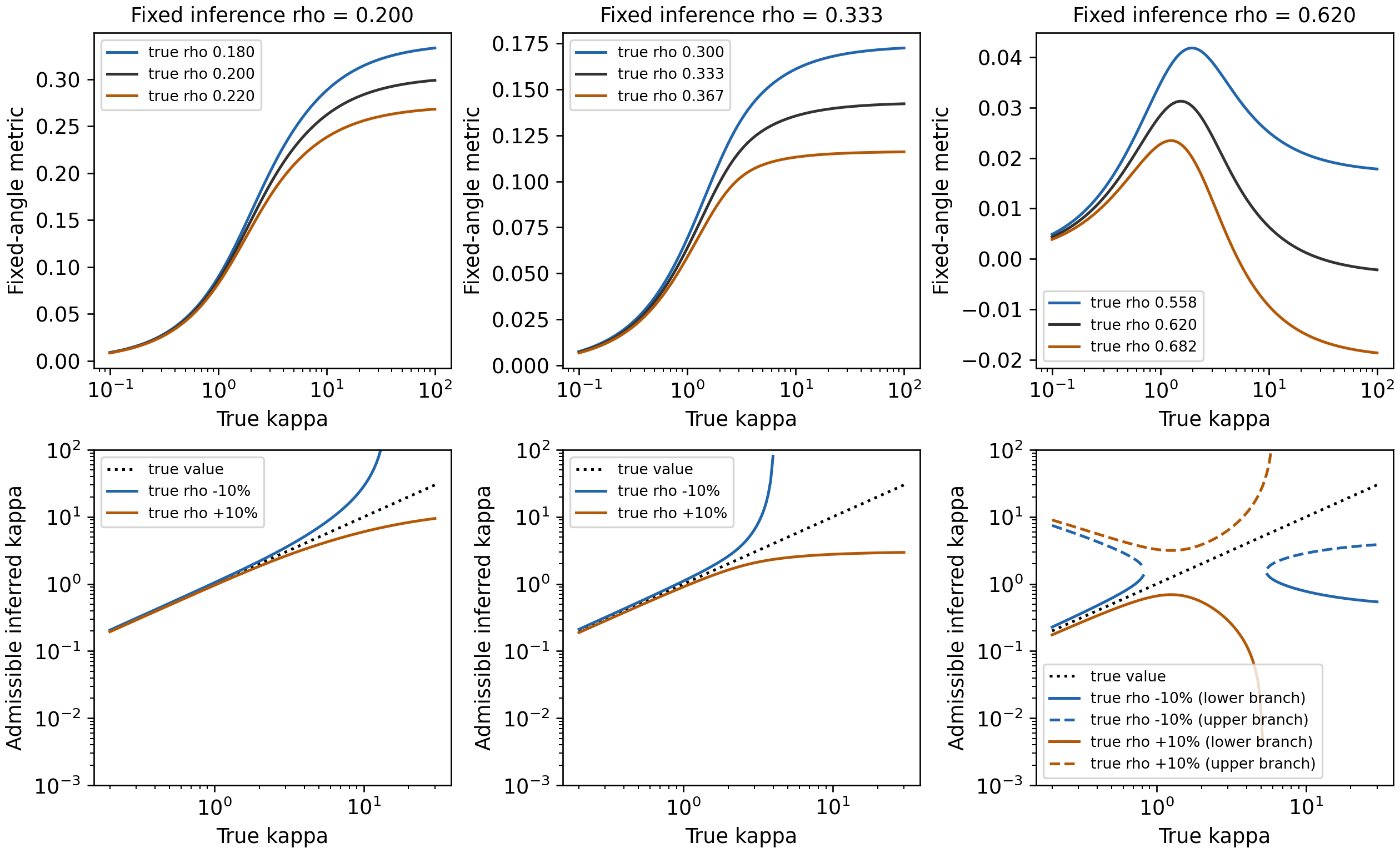}
\caption{Sensitivity to tensor misspecification. Upper panels show the forward metrics for the nominal and perturbed true ratios. Lower panels show the admissible inverse branches for each true concentration. Dashed colored curves track the upper monotone branch when present; the dotted diagonal is exact recovery. A curve can end when its branch loses a root or leaves the displayed axes. Roots beyond the displayed axes are retained in the data tables. No branch is selected to improve agreement.}
\label{fig:S10rho}
\end{figure}

All solutions in $0\leq\hat\kappa\leq1000$ are reported. Stationary points partition the nominal response into monotone branches; bracketing on each branch prevents selection of only one favorable root. Stationary-point detection is checked using both 2,001 and 4,001 logarithmically spaced points before refinement. A missing root is flagged as no solution in this declared computational range, not a physical upper limit on concentration or proof about roots outside it. Table~\ref{tab:S10rho} and Fig.~\ref{fig:S10rho} show both severe bias and ambiguous inversion. Data tables retain all roots, including those outside displayed plot limits.

\begin{table}[htbp]
\centering
\caption{Admissible inferred $\hat\kappa$ for true $\rho$ 10\% below or above the fixed nominal value. Multiple entries are distinct inverse solutions. An em dash denotes no solution in $[0,1000]$. These are noise-free model examples, not experimental parameter estimates.}
\label{tab:S10rho}
\begin{tabular}{rrrr}
\hline
Assumed $\rho$ & True $\kappa$ & True $\rho$: $-10\%$ & True $\rho$: $+10\%$\\
\hline
0.2000 & 2 & 2.183 & 1.839 \\
0.2000 & 5 & 6.999 & 3.876 \\
0.2000 & 20 & \textemdash & 8.295 \\
0.3333 & 2 & 2.607 & 1.621 \\
0.3333 & 5 & \textemdash & 2.481 \\
0.3333 & 20 & \textemdash & 2.917 \\
0.6200 & 2 & \textemdash & 0.544, 3.855 \\
0.6200 & 5 & \textemdash & 0.016, 25.063 \\
0.6200 & 20 & 0.595, 3.586 & \textemdash \\
\hline
\end{tabular}

\end{table}

\subsection{Background Fraction and Reconstructed Trajectories}
\label{sec:S10background}

Let $q=B/S_{45}$ denote the instantaneous common background-to-signal ratio at the denominator angle. Ignoring that background multiplies the true normalized metric by $1/(1+q)$ exactly. For example, at $\rho=0.2$ and true $\kappa=5$, $q=0.01,0.1,0.25,0.5,1$ yield inferred concentrations of approximately $4.832,3.747,2.800,2.041,1.382$. For true $\kappa=20$, even $q=0.01$ gives approximately 17.609. A background fraction is not a relative concentration-error bound, particularly near saturation. Over-subtraction can increase the metric or exceed its admissible range.

\begin{figure}[!htbp]
\centering
\includegraphics[width=\textwidth]{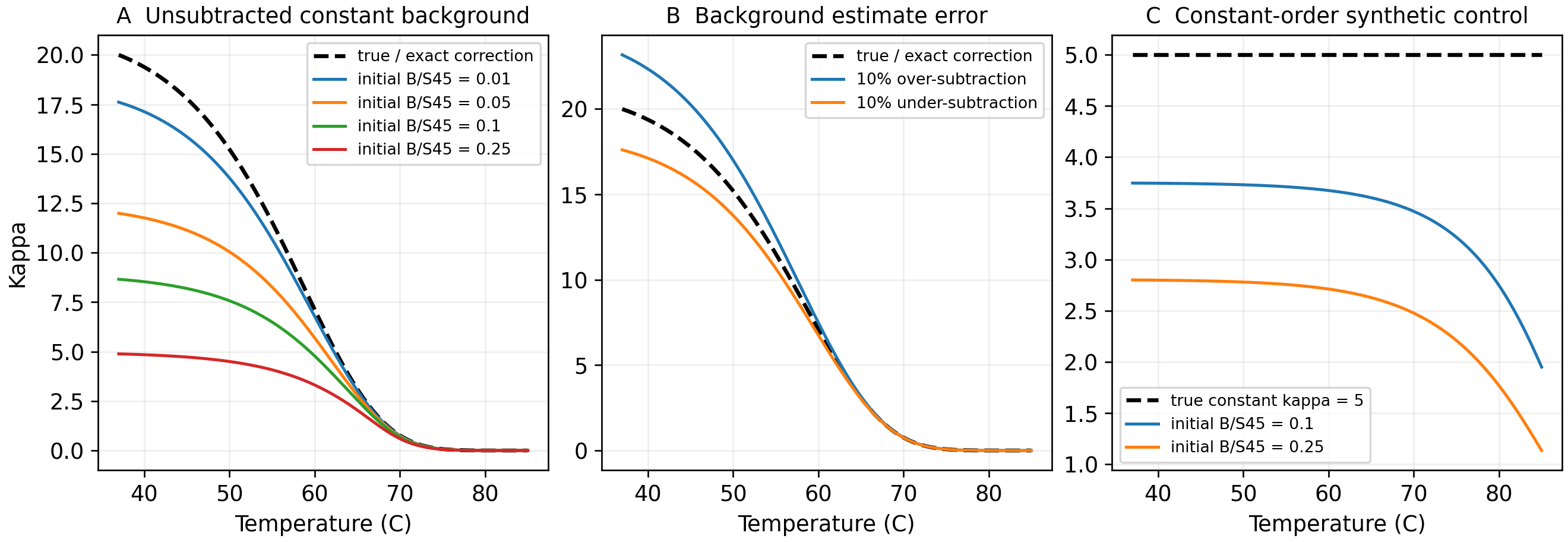}
\caption{Noise-free background sensitivity of reconstructed concentration. (A) Unsubtracted constant backgrounds, labeled by their initial denominator-signal fraction. (B) Exact subtraction and 10\% errors in the estimated physical background at initial fraction 0.1. (C) Constant true concentration with a changing signal scale: an apparent concentration decline occurs without actual orientational change when background is ignored. These illustrative curves are not measured thermal trajectories or predictions of tissue-damage temperatures.}
\label{fig:S10background}
\end{figure}

Figure~\ref{fig:S10background} reuses the existing kinetic equations and parameters, with $\rho=0.2$, $G=1$, a 1$^\circ$C/min ramp, and a 37--85$^\circ$C grid at 0.05$^\circ$C spacing. The true initial concentration is 20. A constant background is chosen by its initial ratio $q_0=B/[\beta_0^2I_\parallel(45^\circ;\kappa_0,\rho)]$, with $q_0=0,0.01,0.05,0.1,0.25$. As $\beta$ and $\kappa$ evolve, the instantaneous fraction changes. The curves show inversion that ignores the background, exact subtraction, and residual errors equal to $\pm10\%$ of the true background when $q_0=0.1$. A separate synthetic control keeps true $\kappa=5$ constant while using the same molecular-response trajectory; its apparent decline isolates a background/scale confound. This control adds no heat-transfer or degradation model.

\clearpage

A deterministic background suppresses the ratio magnitude, whereas stochastic noise adds estimator uncertainty and can bias noisy ratios and threshold crossings. Perfect mean-background subtraction does not remove background photon noise. The existing background-free maximum near $\rho=0.62$ remains a separate intrinsic nonmonotonicity of this fixed-angle metric. Neither its existence nor a noisy peak-and-fall trace is evidence of structural recovery.

\subsection{Verification and Reproduction}

The checks are supplied with their specification, script and full numerical tables in the review-response reproduction folder. Independent periodic quadrature of the local squared field reproduces closed-form intensities to better than $2\times10^{-12}$ in the tested cases. Analytic derivatives are checked by finite differences to absolute tolerance $2\times10^{-8}$, all reported roots reproduce the forward metric to tolerance $2\times10^{-11}$, the exact background factor is verified, and exact subtraction recovers the imposed concentration trajectory. The original model source is checked unchanged. These checks establish numerical consistency under stated assumptions; they do not establish physical parameter identifiability or experimental validity.

\section*{Funding}

This work was supported by the Research Grants Council of Hong Kong Special Administrative Region, China (project \#: CUHK 14210525 and C5005-23WF) and the Hong Kong Innovation and Technology Fund (Grant No. ITS/392/24).

\section*{Acknowledgments}

The authors thank Dr. Shiyang Yuan for assistance with operating the SHG microscope.

\section*{Author Contributions}

Sui Ip Ng and Ken-Tye Yong conceptualized the study. Sui Ip Ng developed the analytical framework and derivations, performed the numerical analysis, experiments and validation, and drafted the manuscript. Yi-Ping Ho provided experimental support and reviewed the manuscript. Ken-Tye Yong supervised the work and reviewed the manuscript.

\bibliography{supplement}